\documentclass[12pt,english,draftcls, one column]{IEEEtran}
\usepackage{amsmath,amsfonts}
\usepackage{algorithmic}
\usepackage{algorithm}
\usepackage{array}
\usepackage[caption=false,font=normalsize,labelfont=sf,textfont=sf]{subfig}
\usepackage{textcomp}
\usepackage{stfloats}
\usepackage{url}
\usepackage{verbatim}
\usepackage{graphicx}
\usepackage{cite}
\usepackage{multirow}
\usepackage{multicol}
\usepackage[T1]{fontenc}
\usepackage[latin9]{inputenc}
\usepackage{float}
\usepackage{amsthm}
\usepackage{amssymb}
\usepackage{xcolor}

\floatstyle{ruled}
\newfloat{algorithm}{tbp}{loa}
\providecommand{\algorithmname}{Algorithm}
\floatname{algorithm}{\protect\algorithmname}

\theoremstyle{plain}
\newtheorem{thm}{\protect\theoremname}
\theoremstyle{plain}
\newtheorem{prop}[thm]{\protect\propositionname}

\usepackage{babel}
\providecommand{\propositionname}{Proposition}
\providecommand{\theoremname}{Theorem}

\begin{document}

\title{\huge Reduced-Complexity WMMSE Approach for Cell-Free Massive MIMO Systems With Finite-Capacity Fronthaul Links}

\author{Wonsik Yoo, \textit{Graduate Student Member}, \textit{IEEE}, Hoon Lee, \textit{Member}, \textit{IEEE}, \\ and Seok-Hwan Park, \textit{Senior Member}, \textit{IEEE} \thanks{
This work was supported in part by the National Research Foundation (NRF) of Korea, funded by the Ministry of Education (MoE) under Grant RS-2025-25423851; in part by the Ministry of Science and ICT (MSIT) under Grants RS-2026-25468472 and RS-2022-NR070834; in part by Korea Research Institute for Defense Technology Planning and Advancement (KRIT) grant funded by the Korean government (DAPA (Defense Acquisition Program Administration)) (21-106-A00-007, Space-Layer Intelligent Communication Network Laboratory, 2022); and 
was supported by the ANCHOR program through the Jeonbuk ANCHOR Center, funded by the Ministry of Education (MOE) and the Jeonbuk State, Republic of Korea (2026-ANCHOR-13-JBU).
\textit{(Corresponding authors: Hoon Lee; Seok-Hwan Park.)}

% Copyright (c) 2026 IEEE. Personal use of this material is permitted. However, permission to use this material for any other purposes must be obtained from the IEEE by sending a request to pubs-permissions@ieee.org. 

W. Yoo is with the Division of Electronic Engineering, Jeonbuk
National University, Jeonju, Korea (email: wonsik0713@jbnu.ac.kr).

H. Lee is with the Department of Electrical Engineering and Artificial Intelligence Graduate School, Ulsan National Institute of Science and Technology (UNIST), Ulsan, Korea (email: hoonlee@unist.ac.kr).

S.-H. Park is with the School of Electrical Engineering, Hanyang University, Ansan, South Korea (email: seokhwanpark@hanyang.ac.kr).
}
}
\maketitle

\begin{abstract}
We propose a reduced-complexity weighted minimum mean squared error (R-WMMSE) framework for cell-free massive MIMO (CF-mMIMO) systems with finite-capacity fronthaul links. To maximize the system capacity, we need to jointly design beamforming vectors at distributed access points (APs) and fronthaul quantization noise covariances. Such a task would be addressed using conventional WMMSE algorithms and their variants. However, existing WMMSE-based methods either assume ideal infinite-capacity fronthaul links or incur high computational complexity when jointly optimizing the network-wide beamforming and fronthaul compression variables. To this end, the proposed method jointly optimizes beamforming and fronthaul compression by leveraging a sequential per-AP optimization strategy. Since each subproblem is efficiently solved via Lagrangian duality with closed-form updates, the overall computational complexity can be significantly reduced compared to the conventional WMMSE method.
To further reduce hardware cost, we extend the R-WMMSE framework to hybrid analog-digital beamforming systems, tackling the joint design of digital beamforming, fronthaul compression, and analog beamforming.
Numerical results show that the proposed scheme achieves performance nearly identical to that of the conventional WMMSE while reducing the runtime by over 99$\%$.
\end{abstract}


\theoremstyle{theorem}
\newtheorem{theorem}{Theorem} 
\theoremstyle{proposition}
\newtheorem{proposition}{Proposition} 
\theoremstyle{lemma}
\newtheorem{lemma}{Lemma} 
\theoremstyle{corollary}
\newtheorem{corollary}{Corollary} 
\theoremstyle{definition}
\newtheorem{definition}{Definition}
\theoremstyle{remark}
\newtheorem{remark}{Remark}

\section{Introduction} \label{sec:intro}

The deployment of cell-free massive multiple-input multiple-output (CF-mMIMO) systems~\cite{Ngo:TWC17, Nayebi:TWC17, Bjornson:TCOM20, Mai:TCOM20, Bjornson:TWC20, Ngo:Proc2024, Zhang:ACC19} has emerged as a promising architecture to provide high spectral efficiency and improved user fairness in next-generation wireless networks.
In the CF-mMIMO systems, a large number of distributed access points (APs) cooperatively serve mobile users under the control of a central processor (CP), thereby eliminating the notion of cell boundaries. Despite their theoretical benefits, CF-mMIMO systems encounter challenges in designing efficient transmission strategies over capacity-constrained fronthaul links. This entails joint optimization of multi-antenna beamforming vectors along with fronthaul compression policies.

Fronthaul-aware downlink design for CF-mMIMO systems has been extensively investigated \cite{NRR:WCL25, Gottsch:TWC24, Lai:TVT25, Mobini:TWC2}. Depending on the open radio access network (O-RAN) functional split and its associated processing and fronthaul requirements, CF-mMIMO systems can be categorized into two cases: Option 7.2 and Option 8.0. Option 7.2 allows each AP to perform PHY functionalities locally. In this case, the CP transmits subsets of data bit streams to the APs via the fronthaul, and the APs perform local baseband processing and beamforming via distributed optimization. However, since each AP can only use locally available information such as local channel state information (CSI), it would fail to suppress multi-user interference. In contrast, under Option 8.0, APs are equipped only with RF sampling and upconversion \cite{NRR:WCL25}, whereas encoding, modulation, and beamforming are carried out at the CP. Thus, the CP conveys compressed beamformed baseband signals to APs through fronthaul links. Such a centralized signal processing strategy efficiently mitigates multi-user interference, thereby enhancing the system capacity. 
Implementing the centralized algorithm at the CP might require fronthaul signaling to collect local CSI from individual APs.

\subsection{Related Works} \label{sub:related-work}

These two functional split options were studied in developing user-centric CF-mMIMO systems \cite{Gottsch:TWC24, Mobini:TWC2, Lai:TVT25}. For functional split Option 7.2, the cluster-wise user-centric beamforming optimization task was addressed in \cite{Mobini:TWC2}.
Due to the strong coupling between beamforming variables and user-AP association variables, closed-form approximations of achievable rates were employed to simplify the optimization problem.
In \cite{Gottsch:TWC24}, both uplink and downlink designs were considered under Option 8.0. Independent compression across the antenna domain was adopted for fronthaul quantization, and the downlink beamforming vectors were chosen as the uplink minimum mean squared error (MMSE) combining vectors.
Similarly, \cite{Lai:TVT25} investigated uplink and downlink designs under the same split option, but quantified the fronthaul load in terms of the number of transmitted complex scalars rather than the actual bit rate. As in \cite{Gottsch:TWC24}, scalable MMSE-based beamforming algorithms were employed.

Overall, these works \cite{Gottsch:TWC24, Mobini:TWC2, Lai:TVT25} primarily emphasize scalable beamforming strategies to ensure practical implementability. However, there remains a need for efficient fronthaul-aware beamforming optimization algorithms that can approach the performance of computationally intensive solutions, such as the weighted minimum mean square error (WMMSE) method \cite{Yu:WCL19}.

In \cite{Liu:TWC25}, a neural network-based unfolding technique was proposed to accelerate the WMMSE algorithm for CF-mMIMO systems. 
Also, for functional split Option 7.2, distributed signal processing techniques were studied in \cite{Zafari:Asilomar24, Wang:TWC25} where individual APs determine their own local beamforming vectors based on the alternating direction method of multipliers algorithm. To realize this method, APs should exchange interaction messages through fronthaul links iteratively. Compared with Option 8.0, which only requires one-shot fronthaul coordination for collecting local CSI, the distributed signal processing in Option 7.2 requires a number of fronthaul usages until convergence. Such signaling overhead becomes prohibitive in large-scale systems.

\subsection{Motivations}

This paper presents fronthaul-aware beamforming optimization techniques for CF-mMIMO systems under Option 8.0. The centralized PHY functionalities at the CP enable efficient beamforming solutions at the expense of a slight increase in fronthaul overhead for CSI collection from APs. Despite such benefits, centralized signal processing algorithms encounter several practical challenges.
One critical issue is the computational complexity of jointly optimizing beamforming vectors across all APs, particularly when leveraging iterative algorithms such as the WMMSE framework \cite{Christensen:TWC08, Yu:WCL19}. While the WMMSE algorithm guarantees convergence to a stationary point, its complexity scales poorly with the number of antennas and APs, limiting its feasibility in large-scale systems.

To address this limitation, recent studies have proposed reduced-WMMSE (R-WMMSE) frameworks \cite{Zhao:TSP23, Yoo:WCL24}.
This approach was first proposed in \cite{Zhao:TSP23} for a multi-user MIMO downlink system under a sum-power constraint (SPC) and a per-antenna power constraint (PAPC).
In the case of PAPC, the beamforming coefficients corresponding to different antennas were sequentially optimized, allowing for closed-form updates without relying on general-purpose optimization solvers such as CVX.
This method has been extended to CF-mMIMO systems in \cite{Yoo:WCL24}.
Since each AP typically uses multiple antennas, per-group-of-antennas (i.e., per-AP) power constraints were considered, and a sequential per-AP optimization strategy was adopted in place of the per-antenna method in \cite{Zhao:TSP23}.
However, \cite{Yoo:WCL24} assumed ideal infinite-capacity fronthaul connections between the APs and the CP, and thus it cannot be applied to practical CF-mMIMO with finite-capacity fronthaul links.

Given that fronthaul capacity limitations are a critical bottleneck in practical CF-mMIMO deployments \cite{Park:TSP13, Yu:WCL19}, this work extends the R-WMMSE framework \cite{Zhao:TSP23, Yoo:WCL24} to systems with finite-capacity fronthaul links.
In such systems, the beamformed signals at the CP need to be quantized and compressed before transmission to the APs,
necessitating joint design of beamforming and fronthaul compression strategies \cite{Park:TSP13, Yu:WCL19}.
This joint optimization problem becomes more challenging than the ideal fronthaul case due to the enlarged optimization space and the non-convexity of additional fronthaul capacity constraints.
Such a task might be addressed by using the modified WMMSE algorithm proposed in \cite{Yu:WCL19} for CF-mMIMO with capacity-constrained fronthaul links. However, since it resorts to convex optimization software, its computational complexity becomes prohibitive, especially for large-scale systems. Therefore, it is challenging to implement the conventional WMMSE approach \cite{Shi:TSP11} in practical CF-mMIMO systems with numerous APs.

\subsection{Contributions} \label{sub:main-contribution}

This paper extends the recently developed R-WMMSE methodology \cite{Zhao:TSP23, Yoo:WCL24} to CF-mMIMO systems with capacity-constrained fronthaul links. With careful investigation, the original joint optimization task across all APs can be decomposed into subproblems, each of which is dedicated to optimizing the local beamforming vectors and the quantization noise covariance matrix of an individual AP. Compared to the conventional method \cite{Yu:WCL19} that optimizes all APs simultaneously, our decomposition approach remarkably reduces the dimension of the solution space, enabling a low-complexity implementation.

Due to the coupled beamforming and quantization noise covariance variables, the subproblem requires commercial optimization software, which might entail high computational complexity. To tackle this issue, we employ the Lagrangian duality method. The proposed algorithm addresses the Lagrangian dual problem \cite[Ch. 5]{Boyd:Cambridge04} via a primal-dual subgradient method \cite{Nedic:JOTA09} with closed-form primal and dual updates. According to our duality analysis, for given dual variables, both the local beamforming vectors and quantization noise covariance matrix can be obtained in closed form. Consequently, the primal update step can be realized with simple calculations, resulting in a lightweight implementation.

While the proposed R-WMMSE framework is well suited for large-scale antenna systems, its implementation remains challenging due to the hardware cost associated with a large number of radio frequency (RF) chains.
To mitigate this issue, we further extend the proposed R-WMMSE framework to hybrid analog-digital beamforming systems \cite{Molisch:COMMA17, Sohrabi:JSTSP16, Liang:WCL14, Kim:TCOM19}, where each AP employs significantly fewer RF chains than antennas.
The problem of jointly optimizing the digital beamforming, fronthaul compression, and analog beamforming becomes more difficult due to the multiplicative coupling between analog and digital processing variables, as well as non-convex constant modulus constraints imposed on analog beamforming coefficients.
We tackle this challenging problem by employing a WMMSE-based reformulation and an alternating optimization (AO) algorithm that iteratively updates the analog and digital processing variables.

Our main contributions are summarized as follows:
\begin{itemize}
    \item We extend the R-WMMSE framework \cite{Zhao:TSP23, Yoo:WCL24} to CF-mMIMO systems with finite-capacity fronthaul links by enabling joint optimization of beamforming vectors and quantization noise covariance matrices under fronthaul capacity constraints.
    The optimization is performed in a per-AP sequential manner, and each subproblem is efficiently solved using Lagrangian duality.
    
    \item We further generalize the R-WMMSE framework to hybrid analog-digital beamforming systems \cite{Molisch:COMMA17, Sohrabi:JSTSP16, Liang:WCL14, Kim:TCOM19}. An AO algorithm is employed, where the analog beamforming and digital processing variables are updated in an iterative fashion. When optimizing the analog beamforming matrices for fixed digital variables, we use a gradient descent (GD) approach \cite{Ruder:arX:16,An:ICC23}.
    
    \item Extensive numerical results demonstrate that the proposed R-WMMSE scheme achieves performance nearly identical to that of the conventional WMMSE scheme while reducing the runtime by over 99$\%$. Furthermore, its hybrid extension achieves performance approaching the fully-digital bound with significantly fewer RF chains, validating its practical efficiency.
\end{itemize}

\subsection{Paper Organization and Notations}

Sec. \ref{sec:system_model} describes the system model, including network architecture, CSI assumption, and signal models for digital beamforming and fronthaul compression. Sec. \ref{sec:problem_formulation} formulates the weighted sum-rate maximization problem under per-AP power and fronthaul capacity constraints and reviews the conventional WMMSE approach.
In Sec. \ref{sec:proposed_fullyd}, we present the proposed R-WMMSE framework, which is then extended to the hybrid analog-digital beamforming systems in Sec. \ref{sec:hybrid_extension}.
Sec. \ref{sec:results} provides numerical results validating the effectiveness of the proposed R-WMMSE schemes. Lastly, Sec. \ref{sec:conclusion} concludes the paper.

The superscripts $(\cdot)^T$, $(\cdot)^H$, and $(\cdot)^*$ denote transpose, Hermitian transpose, and complex conjugate, respectively. $\mathbb{E}[\cdot]$ denotes expectation, and $\operatorname{tr}(\cdot)$ and $\det(\cdot)$ denote the trace and determinant of a matrix.
The operators $\mathrm{diag}(\cdot)$ and $\mathrm{blkdiag}(\cdot)$ denote diagonal and block-diagonal matrices. $\mathcal{CN}(\mu, \Sigma)$ denotes a circularly symmetric complex Gaussian distribution with mean $\mu$ and covariance $\Sigma$. The identity matrix of size $d$ is denoted by $\mathbf{I}_d$, and $\|\cdot\|$ represents the Euclidean norm.

\section{System Model} \label{sec:system_model}

\begin{figure}
\centering\includegraphics[width=0.6\linewidth]{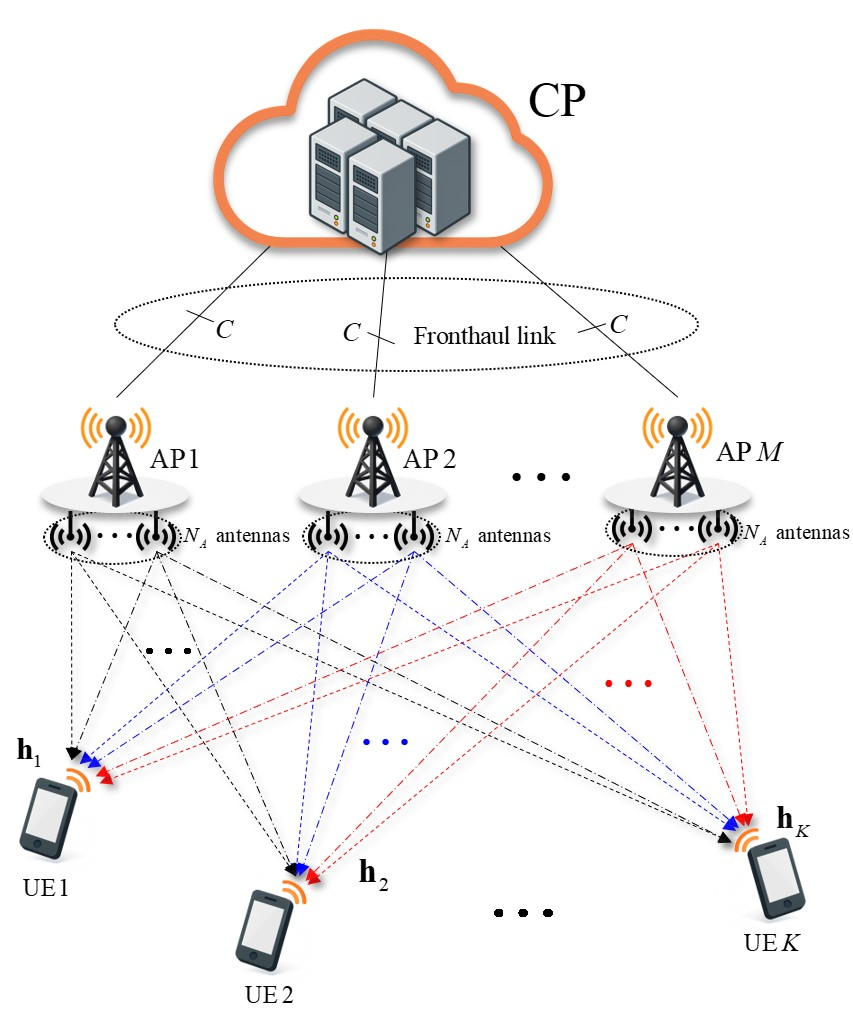}
%\vspace{-5mm}
\caption{\label{fig:system-model}Illustration of CF-mMIMO downlink system}
\end{figure}

As illustrated in Fig.~\ref{fig:system-model}, we consider the downlink of a CF-mMIMO system comprising $M$ distributed APs where each AP $i$ with $N_A$ antennas is connected to a CP via a fronthaul link of finite capacity $C_i$ bps/Hz.
The system serves $K$ single-antenna user equipments (UEs), indexed by $\mathcal{K} = \{1, \dots, K\}$, through the APs indexed by $\mathcal{M} = \{1, \dots, M\}$.
The time-division duplexing (TDD) mode is adopted where individual APs can acquire downlink channels to the UEs based on uplink channel training procedures \cite{Nayebi:TWC17, Shaik:TCOM21}. With estimated CSI, the APs apply multi-antenna signal processing to the data symbols sent by the CP to maximize the overall system performance.
Throughout the paper, except in Sec. \ref{sec:hybrid_extension}, we assume fully-digital processing, i.e., $N_R = N_A$, where $N_R$ denotes the number of RF chains at each AP.
In Sec. \ref{sec:hybrid_extension}, we address the case with limited RF chains, i.e., $N_R < N_A$, where analog beamforming between the RF chains and antennas needs to be jointly designed with digital beamforming.
Throughout the paper, we assume that the synchronization among APs and between APs and UEs is perfect\footnote{The assumption of perfect synchronization is particularly crucial in large-scale networks, where phase-calibration may become unbounded \cite{Larsson:TSP24}. Developing robust algorithms that account for imperfect synchronization is beyond the scope of this work and is left for future research.}.

\subsection{Channel Acquisition} \label{sub:channel_model}

Let $\mathbf{h}_{k,i}\sim\mathcal{CN}(\mathbf{0}, \mathbf{R}_{k,i})\in\mathbb{C}^{N_A\times 1}$ denote the Rayleigh fading channel vector between UE $k$ and AP $i$, where $\mathbf{R}_{k,i}$ is the covariance matrix of $\mathbf{h}_{k,i}$. To acquire the channel vectors at the APs, the UEs convey $L$ orthogonal pilot sequences, each of length $L$. Due to the limited time resources, the pilot length $L$ is typically smaller than the number of UEs $K$, thereby posing the pilot contamination issue. Let $l_{k}$ be the index of the pilot signal ($l_{k} \in \{1,\cdots,L\}$) selected by UE $k$. Also, the set of UEs assigned to the $l$th pilot is denoted by $\mathcal{K}_{l} \subset \mathcal{K}$.

Based on the received pilot signals, each AP $i$ obtains an estimate $\{\widehat{\mathbf{h}}_{k,i}\}_{k\in\mathcal{K}}$ of its local CSI $\{\mathbf{h}_{k,i}\}_{k\in\mathcal{K}}$ using a linear MMSE estimator \cite{Choi:TWC20, Nayebi:TWC17, Shaik:TCOM21}.
The estimated channel $\widehat{\mathbf{h}}_{k,i}$ is then related to the actual channel vector $\mathbf{h}_{k,i}$ as
\begin{align}
    \mathbf{h}_{k,i} = \widehat{\mathbf{h}}_{k,i} + \mathbf{e}_{k,i}, 
    \label{eq:moedl_CSI_error_ki}
\end{align}
where $\mathbf{e}_{k,i}\sim\mathcal{CN}(\mathbf{0}, \mathbf{E}_{k,i})$ is the estimation error uncorrelated with $\widehat{\mathbf{h}}_{k,i}$.
The error covariance matrix $\mathbf{E}_{k,i}$ is given by $\mathbf{E}_{k,i} = \mathbf{R}_{k,i} - \widehat{\mathbf{R}}_{k,i}$ \cite{Shaik:TCOM21, Choi:TWC20}, where $\widehat{\mathbf{R}}_{k,i}$ denotes the covariance matrix of $\widehat{\mathbf{h}}_{k,i}$ given by 
\begin{align}
    \widehat{\mathbf{R}}_{k,i} = \mathbf{R}_{k,i}\left( \sum\nolimits_{l\in\mathcal{K}_{l_k}} \mathbf{R}_{l,i} + \frac{N_0^{\text{ul}}}{L P^{\text{ul}}}\mathbf{I}_{N_A}\right)^{-1}\mathbf{R}_{k,i}. \label{eq:nominal-CSI-variance}
    %\frac{ \rho_{k,i}^2 }{ 1/\left(L\frac{P^{\text{ul}}}{N_0^{\text{ul}}}
    %\right) + \sum_{l\in\mathcal{K}_{l_k}} \rho_{l,i}}. \label{eq:nominal-CSI-variance}
\end{align}
Here, $P^{\text{ul}}$ and $N_0^{\text{ul}}$ denote the uplink transmission power of the UEs and the noise variance at the APs, respectively.
For notational convenience, we define stacked vectors ${\mathbf{h}}_{k}=[{\mathbf{h}}_{k,1}^H \, \cdots \, \mathbf{h}_{k,M}^H]^H$ and $\mathbf{e}_k=[\mathbf{e}_{k,1}^H \, \cdots \, \mathbf{e}_{k,M}^H]^H\sim\mathcal{CN}(\mathbf{0},\mathbf{E}_k)$, where $\mathbf{E}_k=\text{blkdiag}(\mathbf{E}_{k,1},\cdots,\mathbf{E}_{k,M})$.
We assume that the CP has access to the estimated channel vectors $\{\widehat{\mathbf{h}}_k\}_{k\in\mathcal{K}}$, reported by the APs, and the corresponding error covariance matrices $\{\mathbf{E}_k\}_{k\in\mathcal{K}}$\footnote{The fronthaul overhead for reporting the estimated channel vectors and error covariance matrices may become prohibitive, particularly for large $N_A$. To address this issue, we will consider extending the model-based neural fronthaul coordination framework developed in \cite{Yu:TC25} to our scenario as a direction for future work.}.

\subsection{Digital Beamforming and Fronthaul Compression} \label{sub:sys_beamforming}

The CP generates a data symbol $s_k \sim \mathcal{CN}(0,1)$ for each UE $k \in \mathcal{K}$ and applies digital beamforming followed by fronthaul quantization and compression.
We adopt the Gaussian test-channel model \cite{Park:TSP13} to characterize fronthaul compression. This model represents the compression process by adding Gaussian quantization noise that is independent of the precoded signal, enabling the fronthaul capacity constraint to be expressed in a tractable log-det form.
% By modeling the quantization process using a Gaussian test channel \cite{Park:TSP13}, 
Accordingly, the transmitted signal $\mathbf{x}_i \in \mathbb{C}^{N_A\times 1}$ by AP $i$ is given by
\begin{align}
    \mathbf{x}_i = \sum_{k \in \mathcal{K}} \mathbf{v}_{k,i} s_k + \mathbf{q}_i,    
    \label{eq:fully_beamforming}
\end{align}
where $\mathbf{v}_{k,i}\in\mathbb{C}^{N_A\times1}$ is the beamforming vector for UE $k$ associated with AP $i$, and $\mathbf{q}_i \sim \mathcal{CN}(\mathbf{0}, \boldsymbol{\Omega}_i)$ represents the quantization noise caused by fronthaul compression.
According to a standard result from source coding theory \cite[Ch. 3]{Gamal:Cambridge11}, the quantized signal $\mathbf{x}_i$ can be reliably decompressed at AP $i$ if the condition $I(\sum_{k \in \mathcal{K}} \mathbf{v}_{k,i} s_k; \mathbf{x}_i) \leq C_i$ holds. Under the Gaussian test channel model \cite{Park:TSP13}, this condition reduces to
\begin{align}
    \log_2 \det\left( \sum_{k \in \mathcal{K}} \mathbf{v}_{k,i} \mathbf{v}_{k,i}^H + \boldsymbol{\Omega}_i \right) - \log_2 \det(\boldsymbol{\Omega}_i) \leq C_i.
    \label{eq:fully_fronthaul_constraints}
\end{align}
Additionally, $\mathbf{x}_i$ is subject to a transmit power constraint given by
\begin{align}
    \mathbb{E}\left[\|\mathbf{x}_i\|^2\right] = \sum_{k \in \mathcal{K}} \operatorname{tr}\left(\mathbf{v}_{k,i} \mathbf{v}_{k,i}^H\right) + \operatorname{tr}\left(\boldsymbol{\Omega}_i\right) \leq P_i.    
    \label{eq:fully_power_constraints}
\end{align}

Under the transmission model in (\ref{eq:fully_beamforming}), the achievable data rate\footnote{The expression in (\ref{eq:rate_fullyd}) represents the spectral efficiency (SE) in bits per second per Hertz (bps/Hz). The corresponding data rate in bps is obtained simply by multiplying the SE by the system bandwidth. Since the bandwidth is a constant factor, it does not affect the optimization problem and is therefore omitted throughout the paper. However, in the numerical results, the reported data rates are presented in bps by multiplying the SE by the bandwidth.} $R_k$ at UE $k$ is given by
\begin{align}
    R_k = \log_2\left(1 + \frac{\big|\widehat{\mathbf{h}}_k^H \mathbf{v}_k\big|^2}{\text{IF}_k(\mathbf{v}, \boldsymbol{\Omega})}\right),
    \label{eq:rate_fullyd}
\end{align}
where $\mathbf{v}_k=[\mathbf{v}_{k,1}^H, \mathbf{v}_{k,2}^H, \cdots, \mathbf{v}_{k,M}^H]^H$ is the aggregate beamforming vector for UE $k$ and the interference-plus-noise power $\text{IF}_k(\mathbf{v}, \boldsymbol{\Omega})$ is defined as
\begin{align}
    \text{IF}_k(\mathbf{v}, \boldsymbol{\Omega}) &= \sum_{l \in \mathcal{K}\setminus \{k\}} \left|\widehat{\mathbf{h}}_k^H \mathbf{v}_l\right|^2 + \sum_{l \in \mathcal{K}} \mathbf{v}_l^H \mathbf{E}_k \mathbf{v}_l \nonumber \\ &+ \widehat{\mathbf{h}}_k^H \bar{\boldsymbol{\Omega}} \widehat{\mathbf{h}}_k + \operatorname{tr}\left(\mathbf{E}_k \bar{\boldsymbol{\Omega}}\right) + N_0^{\text{dl}},
    \label{eq:IF_func_fullyd}
\end{align}
with $\bar{\boldsymbol{\Omega}}=\text{blkdiag}(\boldsymbol{\Omega}_1, \dots, \boldsymbol{\Omega}_M)$, $\mathbf{v}=\{\mathbf{v}_{k,i}\}_{k\in\mathcal{K},\, i\in\mathcal{M}}$, and $\boldsymbol{\Omega}=\{\boldsymbol{\Omega}_i\}_{i\in\mathcal{M}}$.
Here, $N_0^{\text{dl}}$ accounts for the noise variance at the UE in the downlink.

\section{Problem Formulation and Conventional WMMSE} \label{sec:problem_formulation}
In this paper, we aim at jointly optimizing the beamforming vectors $\mathbf{v}$ and the quantization noise covariance matrices $\boldsymbol{\Omega}$ to maximize the weighted sum-rate performance under the fronthaul capacity constraint \eqref{eq:fully_fronthaul_constraints} and the power constraint \eqref{eq:fully_power_constraints}. Denoting $\alpha_{k}>0$ as a weight assigned to UE $k$, the weighted sum-rate maximization task can be stated as
\begin{subequations} \label{prob:original_fullyd}
    \begin{align} 
        \underset{\mathbf{v}, \boldsymbol{\Omega}}{\max}& \:\: \sum_{k \in \mathcal{K}} \alpha_k R_k  \label{prob:obj_fullyd} \\
        \mathrm{s.t.}&\:\: p_i\left(\mathbf{v},\boldsymbol{\Omega}\right) \leq P_i, \: \forall i \in \mathcal{M}, \label{prob:tpc_fullyd}\\
        &\:\: \log_2 \det\left( \sum_{k \in \mathcal{K}} \mathbf{v}_{k,i} \mathbf{v}_{k,i}^H + \boldsymbol{\Omega}_i \right) \notag \\ 
        &\quad\quad\quad - \log_2 \det(\boldsymbol{\Omega}_i) \leq C_i, \: \forall i \in \mathcal{M},\,\label{prob:fc_fullyd}\\
        &\:\: \boldsymbol{\Omega}_i \succeq \mathbf{0}, \: \forall i \in \mathcal{M}, \label{prob:psd_fullyd}
    \end{align}
\end{subequations}
where $p_i(\mathbf{v},\boldsymbol{\Omega}) = \sum_{k \in \mathcal{K}} \operatorname{tr}(\mathbf{v}_{k,i} \mathbf{v}_{k,i}^H) + \operatorname{tr}(\boldsymbol{\Omega}_i)$ indicates the transmit power at AP $i$.
By appropriately selecting the weights $\{\alpha_k\}_{k\in\mathcal{K}}$, problem (\ref{prob:original_fullyd}) enables flexible fairness control among UEs. For instance, the proportional fairness criterion in \cite{Viswanath:TIT02} assigns larger weights to UEs experiencing lower historical rates.
Problem (\ref{prob:original_fullyd}) is non-convex due to the objective function in (\ref{prob:obj_fullyd}) and the fronthaul capacity constraints (\ref{prob:fc_fullyd}).

To address the non-convexity, reference \cite{Yu:WCL19} extended the WMMSE method \cite{Christensen:TWC08} to the considered CF-mMIMO system with finite-capacity fronthaul.
Specifically, using the WMMSE transformation \cite{Christensen:TWC08} and Fenchel's inequality \cite[Lem. 1]{Zhou:TSP16}, problem \eqref{prob:original_fullyd} can be equivalently reformulated as
\begin{subequations} \label{prob:WMMSE_fullyd}
    \begin{align}
    \underset{{\mathbf{v}, \boldsymbol{\Omega}, \mathbf{u}, \mathbf{w},\boldsymbol{\Sigma}}}{\min} & \:\:\:\, \sum_{k \in \mathcal{K}} \alpha_k \left( w_k \text{MSE}_k(\mathbf{v}, \boldsymbol{\Omega},u_k) - \ln w_k\right) \label{prob:WMMSE_objective_fullyd} \\
    \mathrm{s.t}&.\:\: p_i\left(\mathbf{v},\boldsymbol{\Omega}\right) \leq P_i, \: \forall i \in \mathcal{M},\\
    & \,\,\,\,\ln\det(\boldsymbol{\Sigma}_i) + \operatorname{tr}\left( \boldsymbol{\Sigma}_i^{-1} \left(\sum_{k\in\mathcal{K}} \mathbf{v}_{k,i} \mathbf{v}_{k,i}^H + \boldsymbol{\Omega}_i\right)\right) \nonumber \\ & \,\,\,\,\,\,\,- N_A - \ln\det(\boldsymbol{\Omega}_i) \leq C_i \ln 2,\:\forall i\in\mathcal{M}, \label{eq:WMMSE_fullyd_fronthaul_constraint}\\
    &\quad \boldsymbol{\Omega}_i \succeq \mathbf{0}, \: \boldsymbol{\Sigma}_i \succeq \mathbf{0},\:\forall i \in \mathcal{M},\\
    &\quad w_k > 0, \:\forall k \in \mathcal{K},
\end{align}    
\end{subequations}
where $\text{MSE}_k(\mathbf{v}, \boldsymbol{\Omega}, u_k)$ denotes the mean squared error (MSE) for UE $k$ with receive filter $u_k\in \mathbb{C}$ given by
\begin{align}
    \text{MSE}_k&(\mathbf{v},\boldsymbol{\Omega},u_k) = \mathbb{E}\left[ |s_k - u_k^* y_k|^2 \right] \nonumber \\ &=\big|1-u_k^* \widehat{\mathbf{h}}_k^H \mathbf{v}_k\big|^2+|u_k|^2\, \text{IF}_k(\mathbf{v},\boldsymbol{\Omega}).
\end{align}
We have also defined the notations $\mathbf{u}=\{u_k\}_{k\in\mathcal{K}}$, $\mathbf{w}=\{w_k\}_{k\in\mathcal{K}}$, and $\boldsymbol{\Sigma}=\{\boldsymbol{\Sigma}_i\}_{i\in\mathcal{M}}$.

For fixed $\{\mathbf{v},\boldsymbol{\Omega}\}$, the optimal $\{\mathbf{u},\mathbf{w},\boldsymbol{\Sigma}\}$  are obtained in closed form as
\begin{subequations} \label{eq:optimal_u_w_sigma}
\begin{align}
    &u_k = \frac{\widehat{\mathbf{h}}_k^H \mathbf{v}_k}{\big|\widehat{\mathbf{h}}_k^H \mathbf{v}_k\big|^2 + \text{IF}_k(\mathbf{v}, \boldsymbol{\Omega})}, \: w_k = \frac{1}{1-u_k^* \widehat{\mathbf{h}}_k^H \mathbf{v}_k}, \label{eq:optimal_u_w} \\
    &\boldsymbol{\Sigma}_i = \sum_{k\in\mathcal{K}} \mathbf{v}_{k,i} \mathbf{v}_{k,i}^H + \boldsymbol{\Omega}_i. \label{eq:optimal_sigma}
\end{align}    
\end{subequations}
For fixed $\{\mathbf{u},\mathbf{w},\boldsymbol{\Sigma}\}$, problem (\ref{prob:WMMSE_fullyd}) is convex in $\{\mathbf{v},\boldsymbol{\Omega}\}$.
Therefore, a local optimum can be found via alternating optimization (AO), which iteratively updates the auxiliary variables $\{\mathbf{u},\mathbf{w},\boldsymbol{\Sigma}\}$ in closed form and optimizes $\{\mathbf{v},\boldsymbol{\Omega}\}$ using a convex optimization solver such as the CVX \cite{Grant:CVX20}. Such AO procedures are repeated until convergence.

\section{Proposed R-WMMSE Approach} \label{sec:proposed_fullyd}

The conventional WMMSE algorithm presented in Sec.~\ref{sec:problem_formulation} leads to a tractable problem, thereby yielding a locally optimal solution. However, since it heavily relies on computationally-expensive convex optimization software, it entails high computational complexity, particularly in large-scale CF-mMIMO systems.

To mitigate this issue, we develop an R-WMMSE algorithm for the joint optimization of beamforming and fronthaul compression by extending the recently developed R-WMMSE method in \cite{Yoo:WCL24}, which was originally designed for CF-mMIMO systems with ideal infinite-capacity fronthaul links. The key idea is to decompose the optimization problem of $\{\mathbf{v},\boldsymbol{\Omega}\}$ for fixed $\{\mathbf{u},\mathbf{w},\boldsymbol{\Sigma}\}$ into per-AP subproblems. Each subproblem is dedicated to the optimization of the beamforming vectors and quantization noise covariance matrix of a specific AP while keeping those of the other APs fixed.
These sequential per-AP updates significantly reduce the overall complexity and allow scalable implementation.

\subsection{Per-AP Optimization} \label{sub:per_AP_problem}

We first establish per-AP optimization tasks which decouple the joint optimization problem in \eqref{prob:WMMSE_fullyd} across individual APs. Let $\mathbf{v}_{A,i} = [\mathbf{v}_{1,i}^H \mathbf{v}_{2,i}^H \cdots \mathbf{v}_{K,i}^H]^H \in \mathbb{C}^{N_A K\times 1}$ denote the stacked beamforming vector associated with AP $i$. We fix the variables $\{\mathbf{v}_{A,j}, \boldsymbol{\Omega}_j\}_{j\in\mathcal{M}\setminus\{i\}}$ for the other APs.
%This subsection focuses on the joint optimization of the beamformer $\mathbf{v}_{A,i}$ and the quantization noise covariance matrix $\boldsymbol{\Omega}_i$ for AP $i$, while keeping the variables $\{\mathbf{v}_{A,j}, \boldsymbol{\Omega}_j\}_{j\in\mathcal{M}\setminus\{i\}}$ for the other APs fixed.
Then, by discarding the terms in the objective function \eqref{prob:WMMSE_objective_fullyd} independent of $\{\mathbf{v}_{A,i}, \boldsymbol{\Omega}_i\}$, the optimization problem at AP $i$ can be formulated as
\begin{subequations} \label{prob:RWMMSE_fullyd}
    \begin{align}
        &\!\!\!\! \underset{\mathbf{v}_{A,i},\boldsymbol{\Omega}_i}{\min}\mathbf{v}_{A,i}^H \mathbf{Q}_i \mathbf{v}_{A,i} + 2 \Re\{\mathbf{b}_{i}^H \mathbf{v}_{A,i}\} + \operatorname{tr}\left(\mathbf{D}_{i} \boldsymbol{\Omega}_i \right)  \label{prob:RWMMSE_objective_fullyd} \\
        &\!\!\!\!\text{s.t.}\:\, \|\mathbf{v}_{A,i}\|^2 + \operatorname{tr}\left(\boldsymbol{\Omega}_i \right) \leq P_i, \label{prob:RWMMSE_tpc_fullyd} \\
        &\:\:\:\: \|\left(\mathbf{I}_K \otimes \mathbf{S}_i\right)\mathbf{v}_{A,i}\|^2 + \operatorname{tr}\left(\boldsymbol{\Sigma}_i^{-1}\boldsymbol{\Omega}_i\right)-\ln{\det \boldsymbol{\Omega}_i} \leq t_i,\label{prob:RWMMSE_fc_fullyd} \\
        &\:\:\:\: \boldsymbol{\Omega}_i \succeq \mathbf{0},
    \end{align}
\end{subequations}
where $\mathbf{Q}_i=\mathbf{I}_K \otimes \mathbf{D}_i \succeq \mathbf{0}$, $\mathbf{S}_i=\boldsymbol{\Sigma}_i^{-1/2}$, $t_i=C_i \ln{2} + N_A -\ln{\det \boldsymbol{\Sigma}_i}$, $\mathbf{D}_i=\widehat{\mathbf{H}}_i \mathbf{U} \widehat{\mathbf{H}}_i^H + \sum_{k\in\mathcal{K}}\alpha_k w_k |u_k|^2 \mathbf{E}_{k,i}$ with $\widehat{\mathbf{H}}_i = [\widehat{\mathbf{h}}_{1,i}, \cdots, \widehat{\mathbf{h}}_{K,i}]$, and $\mathbf{U}=\text{diag}( \{ \alpha_k w_k |u_k|^2\}_{k\in\mathcal{K}})$. Also, $\mathbf{b}_{i}$ in \eqref{prob:RWMMSE_objective_fullyd} is defined as
\begin{align}\label{eq:bi}
    \mathbf{b}_i&=-\mathbf{W}\widehat{\mathbf{h}}_{A,i} + \left(\mathbf{I}_K \otimes \widehat{\mathbf{H}}_i\right)\nonumber\\
    &
    ~~~~~ \times \Bigg(\sum_{m\in\mathcal{M}\setminus\{i\}}\left(\mathbf{I}_K\otimes\mathbf{U} \widehat{\mathbf{H}}_m^H\right) \mathbf{v}_{A,m}\Bigg),
\end{align}
where $\mathbf{W}=\text{blkdiag}( \{ \alpha_k w_k u_k \mathbf{I}_{N_A}\}_{k\in\mathcal{K}})$ and $\widehat{\mathbf{h}}_{A,i} = [\widehat{\mathbf{h}}_{1,i}^H, \cdots, \widehat{\mathbf{h}}_{K,i}^H]^H$.

Since problem (\ref{prob:RWMMSE_fullyd}) is convex and satisfies Slater's condition \cite[Ch. 5]{Boyd:Cambridge04}, the strong duality holds. This allows us to solve (\ref{prob:RWMMSE_fullyd}) optimally using the Lagrangian duality method \cite[Ch. 5]{Boyd:Cambridge04}. Let $\lambda_i \geq 0$ and $\mu_i \geq 0$ be dual variables associated with the power constraint \eqref{prob:RWMMSE_tpc_fullyd} and the fronthaul capacity constraint \eqref{prob:RWMMSE_fc_fullyd}, respectively. Denoting $\boldsymbol{\rho}_i = [\lambda_i \,\, \mu_i]^T$ as the dual variable vector,
the Lagrangian for (\ref{prob:RWMMSE_fullyd}) is given by
\begin{subequations} \label{eq:RWMMSE_lagragian_fullyd}
\begin{align} 
    \mathcal{L}&\left( \mathbf{v}_{A,i}, \boldsymbol{\Omega}_i, \boldsymbol{\rho}_i \right) = \mathbf{v}_{A,i}^H \mathbf{Q}_i \mathbf{v}_{A,i} + 2 \Re\{\mathbf{b}_{i}^H \mathbf{v}_{A,i}\} \\ 
    &+ \operatorname{tr}(\mathbf{D}_{i} \boldsymbol{\Omega}_i ) + \lambda_i \left( \|\mathbf{v}_{A,i}\|^2 + \operatorname{tr}(\boldsymbol{\Omega}_i) - P_i\right) \nonumber \\
    & + \mu_i \left( \|(\mathbf{I}_K \otimes \mathbf{S}_i)\mathbf{v}_{A,i}\|^2 + \operatorname{tr}(\boldsymbol{\Sigma}_i^{-1}\boldsymbol{\Omega}_i)-\ln{\det \boldsymbol{\Omega}_i} - t_i\right) \nonumber\\
    &=\mathcal{L}_{V}( \mathbf{v}_{A,i}, \boldsymbol{\rho}_i)+\mathcal{L}_{\Omega}( \boldsymbol{\Omega}_i, \boldsymbol{\rho}_i)-\lambda_{i}P_{i}-\mu_{i}t_{i},
\end{align}
\end{subequations}
where $\mathcal{L}_{V}( \mathbf{v}_{A,i}, \boldsymbol{\rho}_i)$ and $\mathcal{L}_{\Omega}( \boldsymbol{\Omega}_i, \boldsymbol{\rho}_i)$ are respectively defined as
\begin{subequations}
\begin{align}
    \mathcal{L}_{V}( \mathbf{v}_{A,i}, \boldsymbol{\rho}_i)&=\mathbf{v}_{A,i}^{H}(\mathbf{Q}_{i}+\lambda_{i}\mathbf{I}_{N_{A}K}+\mu_{i}(\mathbf{I}_{K}\otimes\boldsymbol{\Sigma}_i^{-1}))\mathbf{v}_{A,i}\nonumber\\&~~+2\Re\{\mathbf{b}_{i}^{H}\mathbf{v}_{A,i}\}, \label{eq:lagragian_v}\\
    \mathcal{L}_{\Omega}( \boldsymbol{\Omega}_i, \boldsymbol{\rho}_i)&=
    %\operatorname{tr}(( \mathbf{D}_i + \lambda_i \mathbf{I}_{N_A} + \mu_i \boldsymbol{\Sigma}_i^{-1})\boldsymbol{\Omega}_{i})\nonumber\\
    \operatorname{tr}(\mathbf{T}_{i}(\boldsymbol{\rho}_{i})\boldsymbol{\Omega}_{i})-\mu_{i}\ln\det\boldsymbol{\Omega}_{i}, \label{eq:lagragian_Om}
\end{align}
\end{subequations}
where $\mathbf{T}_{i}(\boldsymbol{\rho}_{i})=\mathbf{D}_i + \lambda_i \mathbf{I}_{N_A} + \mu_i \boldsymbol{\Sigma}_i^{-1}$.
The dual function is given as
\begin{subequations}\label{eq:gi}
\begin{align}
g_i( \boldsymbol{\rho}_i )&=\min_{\mathbf{v}_{A,i}, \boldsymbol{\Omega}_i\succ\mathbf{0}} \mathcal{L}\left( \mathbf{v}_{A,i}, \boldsymbol{\Omega}_i, \boldsymbol{\rho}_{i}\right)\\
&=\min_{\mathbf{v}_{A,i}}\mathcal{L}_{V}( \mathbf{v}_{A,i}, \boldsymbol{\rho}_i) +
\min_{\boldsymbol{\Omega}_i\succ\mathbf{0}} \mathcal{L}_{\Omega}( \boldsymbol{\Omega}_i, \boldsymbol{\rho}_i)\nonumber\\
& ~~ -\lambda_{i}P_{i}-\mu_{i}t_{i}.    
\end{align}
\end{subequations}
Then, the dual problem is formulated as
\begin{align}
    \underset{\boldsymbol{\rho}_i\succeq\mathbf{0}}{\max} \:\: g_i( \boldsymbol{\rho}_i). 
    \label{prob:RWMMSE_dual_problem}
\end{align}

Notice that the existing work in \cite{Yoo:WCL24} also employs the Lagrangian duality method to optimize each beamforming vector $\mathbf{v}_{A,i}$ under ideal fronthaul links. However, the dual problem in \eqref{prob:RWMMSE_dual_problem} for our scenario is significantly more challenging due to the enlarged optimization space in both the primal and dual domains.

\subsubsection{Update of the primal variables $\{\mathbf{v}_{A,i}, \boldsymbol{\Omega}_i\}$}
We solve the dual problem \eqref{prob:RWMMSE_dual_problem} using the primal-dual subgradient method~\cite{Nedic:JOTA09}, which iteratively updates the primal variables $\{\mathbf{v}_{A,i}, \boldsymbol{\Omega}_i\}$ and the dual variables $\{\lambda_i, \mu_i\}$. We first optimize the primal variables $\{\mathbf{v}_{A,i}, \boldsymbol{\Omega}_i\}$ for given $\boldsymbol{\rho}_{i}$. The beamforming vector $\mathbf{v}_{A,i}$ which minimizes $\mathcal{L}_{V}( \mathbf{v}_{A,i}, \boldsymbol{\rho}_i)$ can be obtained from the stationary condition $\nabla_{\mathbf{v}_{A,i}}\mathcal{L}_{V}( \mathbf{v}_{A,i}, \boldsymbol{\rho}_i)=\mathbf{0}$ as
\begin{align}\label{eq:optimal_solution_beamforming_a}
    \mathbf{v}_{A,i}=-\left( \mathbf{Q}_i + \lambda_i \mathbf{I}_{N_A K} + \mu_i \left( \mathbf{I}_K \otimes \boldsymbol{\Sigma}_i^{-1} \right)\right)^{-1} \mathbf{b}_i.
\end{align}
A direct computation of \eqref{eq:optimal_solution_beamforming_a} incurs $\mathcal{O}(N_{A}^{3}K^{3})$ operations, which mainly stem from the matrix inversion of size $N_{A}K$-by-$N_{A}K$. To reduce this complexity, we further derive $\mathbf{v}_{A,i}$ as
\begin{align}\label{eq:optimal_solution_beamforming_b}
    \mathbf{v}_{A,i}= -\left( \mathbf{I}_K\otimes\mathbf{T}_i(\boldsymbol{\rho}_i)^{-1} \right)\mathbf{b}_{k,i}.
\end{align}
Here, \eqref{eq:optimal_solution_beamforming_b} comes from the facts $\mathbf{Q}_{i}=\mathbf{I}_{K}\otimes\mathbf{D}_{i}$, $\mathbf{I}_{N_{A}K}=\mathbf{I}_{K}\otimes\mathbf{I}_{N_{A}}$, and $(\mathbf{A}\otimes\mathbf{B})^{-1}=\mathbf{A}^{-1}\otimes\mathbf{B}^{-1}$. Since $\mathbf{I}_{K}\otimes\mathbf{T}_{i}(\boldsymbol{\rho}_{i})^{-1}$ is a block diagonal matrix, we can decouple the stacked beamforming vector $\mathbf{v}_{A,i}$ in \eqref{eq:optimal_solution_beamforming_b} across individual $\mathbf{v}_{k,i}$, $\forall k\in\mathcal{K}$, as
\begin{align}
    \mathbf{v}_{k,i}=-\mathbf{T}_{i}(\boldsymbol{\rho}_{i})^{-1}\mathbf{b}_{k,i}, \label{eq:v_k_decoupled}
\end{align}
where $\mathbf{b}_{k,i}\in\mathbb{C}^{N_{A}\times1}$ is subvector $k$ of
$\mathbf{b}_{i}$, i.e.,
$\mathbf{b}_{i}=[\mathbf{b}_{1,i}^{H}\cdots\mathbf{b}_{K,i}^{H}]^{H}$. We can see that \eqref{eq:v_k_decoupled} involves only a single
$N_{A}\times N_{A}$ inversion shared by all $K$ beamforming vectors, thereby
reducing the computational cost to $\mathcal{O}(N_{A}^{3}+KN_{A}^{2})$.

In addition, to determine the noise covariance matrix $\boldsymbol{\Omega}_{i}$, we need to address the minimization task of $\mathcal{L}_{\Omega}( \boldsymbol{\Omega}_i, \boldsymbol{\rho}_i)$, which is expressed as
\begin{align}\label{prob:Omega}
    \min_{\boldsymbol{\Omega}_i\succ\mathbf{0}} \operatorname{tr}(\mathbf{T}_{i}(\boldsymbol{\rho}_{i})\boldsymbol{\Omega}_{i})-\mu_{i}\ln\det\boldsymbol{\Omega}_{i}.
\end{align}
To tackle \eqref{prob:Omega}, we introduce the following proposition which reveals feasibility conditions for the dual variables.
\begin{prop}\label{prop:prop1}
The optimal $\boldsymbol{\rho}_{i}$ for \eqref{prob:RWMMSE_dual_problem} should satisfy 
\begin{align}\label{eq:Ai}
    \mathbf{T}_{i}(\boldsymbol{\rho}_{i})\succ\mathbf{0}.
\end{align}
\end{prop}

\begin{IEEEproof}
We prove \eqref{eq:Ai} by contradiction. First, if $\mathbf{T}_{i}(\boldsymbol{\rho}_{i})=\mathbf{0}$, the solution to \eqref{prob:Omega} can be represented by $\boldsymbol{\Omega}_{i}=a\mathbf{Z}$ for any $a>0$ and $\mathbf{Z}\succ\mathbf{0}$. To minimize $-\mu_{i}\ln\det\boldsymbol{\Omega}_{i}$, $a$ can be set to an arbitrarily large number, which leads to the unbounded optimal value of \eqref{prob:RWMMSE_dual_problem}. Thus, $\mathbf{T}_{i}(\boldsymbol{\rho}_{i})$ cannot be the zero matrix.

Next, suppose that $\mathbf{T}_{i}(\boldsymbol{\rho}_{i})$ has at least one negative eigenvalue. Let the eigenvalue decomposition of $\mathbf{T}_{i}(\boldsymbol{\rho}_{i})$ be $\mathbf{T}_{i}(\boldsymbol{\rho}_{i})=\mathbf{Z}\boldsymbol{\Gamma}\mathbf{Z}^{H}$, where $\mathbf{Z}=[\mathbf{z}_{1},\cdots,\mathbf{z}_{N_{A}}]$ with $\mathbf{z}_{q}$ being the $q$-th eigenvector corresponding to the eigenvalue $\gamma_{q}$ and $\boldsymbol{\Gamma}=\text{diag}(\gamma_{1},\cdots,\gamma_{N_A})$ is the eigenvalue matrix. Without loss of optimality, it is assumed that the first eigenvalue $\gamma_{1}$ is negative. Then, the optimal $\boldsymbol{\Omega}_{i}$ for \eqref{prob:Omega} can be given by
\begin{align}
    \boldsymbol{\Omega}_{i} = a \mathbf{z}_{1}\mathbf{z}_{1}^{H} + \sum_{q=2}^{N_{A}}\epsilon_{q}\mathbf{z}_{q}\mathbf{z}_{q}^{H},
\end{align}
where $a\rightarrow\infty$ and $\epsilon_{q}>0$ is an arbitrary small positive number. In this case, $\mathcal{L}_{\Omega}( \boldsymbol{\Omega}_i, \boldsymbol{\rho}_i)$ in \eqref{eq:lagragian_Om} can be written by
\begin{align}
    \mathcal{L}_{\Omega}( \boldsymbol{\Omega}_i, \boldsymbol{\rho}_i)=a\gamma_{1}+\sum_{q=2}^{N_{A}}\epsilon_{q}\gamma_{q}-\mu_{i}\ln \bigg(a\prod_{q=2}^{N_{A}}\epsilon_{q}\bigg).
\end{align}
It is not difficult to see that $\lim_{a\rightarrow\infty}\mathcal{L}_{\Omega}( \boldsymbol{\Omega}_i, \boldsymbol{\rho}_i)\rightarrow-\infty$, implying that the optimal value for problem \eqref{prob:RWMMSE_dual_problem} becomes unbounded. Combining these results, we can conclude that $\mathbf{T}_{i}(\boldsymbol{\rho}_{i})$ can only have positive eigenvalues, which leads to \eqref{eq:Ai}. This completes the proof.
\end{IEEEproof}

Proposition \ref{prop:prop1} shows that, at the optimum, the matrix $\mathbf{T}_{i}(\boldsymbol{\rho}_{i})$ should be positive definite and invertible. Based on this result, we can attain a closed-form expression of the noise covariance $\boldsymbol{\Omega}_{i}$ that minimizes $\mathcal{L}_{\Omega}(\boldsymbol{\Omega}_{i},\boldsymbol{\rho}_{i})$. From the stationary condition $\nabla_{\boldsymbol{\Omega}_{i}}\mathcal{L}_{\Omega}(\boldsymbol{\Omega}_{i},\boldsymbol{\rho}_{i})=\mathbf{0}$, it follows
\begin{align}\label{eq:L_Omega_0}
    \mu_{i}\boldsymbol{\Omega}_{i}^{-1}=\mathbf{T}_{i}(\boldsymbol{\rho}_{i}),
\end{align}
where we have used the fact $\boldsymbol{\Omega}_{i}\succ\mathbf{0}$. It is revealed from \eqref{eq:L_Omega_0} that the optimal dual variable $\mu_{i}$ should be positive since otherwise $\mathbf{T}_{i}(\boldsymbol{\rho}_{i})=\mathbf{0}$, which contradicts \eqref{eq:Ai}. Consequently, the solution to \eqref{prob:Omega} can be expressed as
\begin{align}\label{eq:optimal_solution_quantization}
    \boldsymbol{\Omega}_i = \mu_i \mathbf{T}_{i}(\boldsymbol{\rho}_{i})^{-1}.
    %\left( \mathbf{D}_i + \lambda_i \mathbf{I}_{N_A} + \mu_i \boldsymbol{\Sigma}_i^{-1} \right)^{-1}.
\end{align}    
We can observe from \eqref{eq:v_k_decoupled} and \eqref{eq:optimal_solution_quantization} that both primal variables $\{\mathbf{v}_{A,i},\boldsymbol{\Omega}_{i}\}$
are determined by the inverse of the same matrix
$\mathbf{T}_{i}(\boldsymbol{\rho}_{i})$. Therefore, \eqref{eq:optimal_solution_quantization} does not require additional computational complexity.

\subsubsection{Update of the dual variables $\{\lambda_i, \mu_i\}$}

Next, we address the dual problem \eqref{prob:RWMMSE_dual_problem} for given primal variables $\{\mathbf{v}_{A,i}, \boldsymbol{\Omega}_{i}\}$. Our analysis in Proposition \ref{prop:prop1} reveals that the dual variables $\boldsymbol{\rho}_{i}$ should fulfill \eqref{eq:Ai}, or, equivalently, 
\begin{align}\label{eq:f}
    \lambda_{i}+\lambda_{\min}(\mathbf{D}_{i}+\mu_{i}\boldsymbol{\Sigma}_{i}^{-1})>0,
\end{align}
with $\lambda_{\min}(\mathbf{Z})$ being the minimum eigenvalue of a matrix $\mathbf{Z}$. 
Since $\mathbf{D}_{i}\succeq\mathbf{0}$ and $\mu_i\boldsymbol{\Sigma}_i^{-1}\succ\mathbf{0}$ for $\mu_i>0$, \eqref{eq:Ai} is guaranteed for all $\lambda_i\geq0$.
Thus, the dual problem \eqref{prob:RWMMSE_dual_problem} can be reformulated as
\begin{subequations}\label{prob:RWMMSE_dual_problem2}
\begin{align}
    \underset{\boldsymbol{\rho}_i}{\max} &\:\: g_i( \boldsymbol{\rho}_i)\\
    \text{s.t.}&\:\: \mu_{i}>0,\\
    &\:\: \lambda_{i} \geq 0.    
\end{align}
\end{subequations}

To tackle \eqref{prob:RWMMSE_dual_problem2}, we adopt the ellipsoid method \cite[Ch. 14]{Boyd:PH91}.
This method iteratively updates an ellipsoidal search region for the dual variable vector $\boldsymbol{\rho}_i$ by shifting and contracting the ellipsoid, until its volume becomes sufficiently small.
At the $t$th iteration, the ellipsoid is characterized by its center vector $\boldsymbol{\rho}_i^{(t)} \in \mathbb{R}^{2\times 1}$ and the volume matrix $\boldsymbol{\Psi}_i^{(t)} \succ \mathbf{0}$.
These are updated from their values at the previous iteration as
\begin{subequations} \label{eq:ellipsoid_update}
    \begin{align}
        \boldsymbol{\rho}_i^{(t)} &\leftarrow \boldsymbol{\rho}_i^{(t-1)} + \frac{1}{3} \boldsymbol{\Psi}_i^{(t-1)} \tilde{\mathbf{m}}_i^{(t-1)}, \label{eq:ellipsoid_center_update} \\
        \boldsymbol{\Psi}_i^{(t)} &\leftarrow \frac{4}{3} \left( \boldsymbol{\Psi}_i^{(t-1)} - \frac{2}{3} \boldsymbol{\Psi}_i^{(t-1)} \tilde{\mathbf{m}}_i^{(t-1)} \left( \tilde{\mathbf{m}}_i^{(t-1)} \right)^T\boldsymbol{\Psi}_i^{(t-1)} \right), \label{eq:ellipsoid_volume_matrix_update}
    \end{align}
\end{subequations}
where $\tilde{\mathbf{m}}_i^{(t)}$ denotes the normalized subgradient direction at the $t$th iteration, which is given as
\begin{align}
    \tilde{\mathbf{m}}_i^{(t)} = \frac{\mathbf{m}^{(t)}}{\sqrt{ (\mathbf{m}^{(t)})^{T} \boldsymbol{\Psi}_i^{(t)} \mathbf{m}^{(t)}}}, \label{eq:ellipsoid_normalized_gradient}
\end{align}
with $\mathbf{m}^{(t)}$ equal to the subgradient vector.

In the ellipsoid method, we calculate the subgradient vector $\mathbf{m}^{(t)}$ according to the feasibility of the current dual variables $\boldsymbol{\rho}_{i}^{(t)}$. If $\boldsymbol{\rho}_{i}^{(t)}$ is feasible for \eqref{prob:RWMMSE_dual_problem2}, $\mathbf{m}^{(t)}$ is set to the subgradient of the dual function $g_{i}(\boldsymbol{\rho}_{i}^{(t)})$ as
\begin{align} \label{eq:m_feasible}
    \mathbf{m}^{(t)} = \left[ \partial_{\lambda_{i}} g_i(\boldsymbol{\rho_{i}}^{(t)})\,\, \partial_{\mu_{i}} g_i(\boldsymbol{\rho_{i}}^{(t)})\right]^T,
\end{align}
where $\partial_{z}$ stands for the subgradient operator with respect to $z$ and the elements of $\mathbf{m}^{(t)}$ can be computed from the constraint violations of the primal problem as
\begin{subequations} \label{eq:subgradient}
    \begin{align}
         \partial_{\lambda_{i}} g_i(\boldsymbol{\rho_{i}}) &= \|\mathbf{v}_{A,i}\|^2 + \operatorname{tr}(\boldsymbol{\Omega}_i) - P_i, \label{eq:subgradient_lambda} \\
         \partial_{\mu_{i}} g_i(\boldsymbol{\rho_{i}}) &= \|(\mathbf{I}_K \otimes \mathbf{S}_i)\mathbf{v}_{A,i}\|^2 + \operatorname{tr}(\boldsymbol{\Sigma}_i^{-1}\boldsymbol{\Omega}_i) \nonumber \\ & \quad\: -\ln{\det \boldsymbol{\Omega}_i} - t_i. \label{eq:subgradient_mu}    
    \end{align}    
\end{subequations}
When it comes to an infeasible $\boldsymbol{\rho}_{i}$, the subgradient vector $\mathbf{m}^{(t)}$ is updated according to the associated infeasible constraint functions. If $\lambda_{i}^{(t)}<0$ or $\mu_{i}^{(t)}\leq0$, it follows 
\begin{align}\label{eq:m_infeasible}
    \mathbf{m}^{(t)}=[1,0]^{T}\ \text{or } \mathbf{m}^{(t)}=[0,1]^{T}.
\end{align}

\subsection{Proposed R-WMMSE Algorithm} \label{sub:RWMMSE_algorithm}

\begin{algorithm}[t]
\caption{Proposed R-WMMSE Algorithm}
\label{algorithm1}
\begin{algorithmic}[1]

\STATE \textbf{initialize:}
\STATE Initialize $\{\mathbf{v}, \boldsymbol{\Omega}\}$ such that (\ref{prob:tpc_fullyd})--(\ref{prob:psd_fullyd}) are satisfied.

\REPEAT
    \STATE Update $\{\mathbf{u},\mathbf{w},\boldsymbol{\Sigma}\}$ with (\ref{eq:optimal_u_w_sigma}).
    \STATE Calculate $\mathbf{c}\leftarrow\sum_{m\in\mathcal{M}} (\mathbf{I}_K\otimes\mathbf{U} \widehat{\mathbf{H}}_m^H) \mathbf{v}_{A,m}$.
    
    \FOR{$i \in \mathcal{M}$}
        \STATE Initialize $t \leftarrow 0$, $\{ \boldsymbol{\rho}_i^{(t)}, \boldsymbol{\Psi}_i^{(t)} \}$ and $\boldsymbol{\rho}_{i}^{\text{best}}\leftarrow\boldsymbol{\rho}_{i}^{(t)}$.
        \STATE Set $\tilde{\mathbf{c}} \leftarrow \mathbf{c} - (\mathbf{I}_K\otimes\mathbf{U} \widehat{\mathbf{H}}_i^H)\mathbf{v}_{A,i}$.
        
        \REPEAT
            \STATE Set $t\leftarrow t+1$.
            \STATE Update $\{\mathbf{v}_{A,i}, \boldsymbol{\Omega}_i\}$ using (\ref{eq:v_k_decoupled}) and \eqref{eq:optimal_solution_quantization} with $\boldsymbol{\rho}_i^{(t-1)}$.
            \STATE Calculate $\mathbf{m}^{(t-1)}$ with \eqref{eq:m_feasible}--\eqref{eq:m_infeasible}.
            \STATE Update the ellipsoid $\{ \boldsymbol{\rho}_i^{(t)}, \boldsymbol{\Psi}_i^{(t)}\}$ with (\ref{eq:ellipsoid_update}).
            \STATE Set $\boldsymbol{\rho}_i^{\text{best}} \leftarrow \arg \max_{\boldsymbol{\rho}_i\in \{\boldsymbol{\rho}_i^{(t)}, \boldsymbol{\rho}_i^{\text{best}}\}} g_i\left( \boldsymbol{\rho}_i\right)$.
        \UNTIL{$\det(\boldsymbol{\Psi}_i^{(t)}) \leq \epsilon_{\mathrm{in}}$.}
        
        \STATE Update $\{\mathbf{v}_{A,i}, \boldsymbol{\Omega}_i\}$ using $\boldsymbol{\rho}_i^{\text{best}}$.
        \STATE Update $\mathbf{c} \leftarrow \tilde{\mathbf{c}} + (\mathbf{I}_K\otimes\mathbf{U} \widehat{\mathbf{H}}_i^H)\mathbf{v}_{A,i}$.
    \ENDFOR
\UNTIL{$R_{\mathrm{var}} \leq \epsilon_{\mathrm{out}}$.}
\end{algorithmic}
\end{algorithm}

Algorithm~\ref{algorithm1} summarizes the detailed procedure of the proposed R-WMMSE algorithm that tackles \eqref{prob:WMMSE_fullyd} in an alternating manner. Unlike the conventional WMMSE method \cite{Yu:WCL19} which jointly optimizes the beamforming vectors $\mathbf{v}$ and the quantization noise covariance matrices $\boldsymbol{\Omega}$ for all APs,
the proposed algorithm sequentially updates $\{\mathbf{v}_{A,i}, \boldsymbol{\Omega}_i\}$ for each AP $i\in\mathcal{M}$ in a block-wise manner.

We first initialize the primal variables $\{\mathbf{v},\boldsymbol{\Omega}\}$ such that the constraints (\ref{prob:tpc_fullyd})--(\ref{prob:psd_fullyd}) are satisfied. 
Using the initialized variables, we compute the initial weighted sum-rate as $R_{\mathrm{sum}}^{\mathrm{old}}=\sum_{k\in\mathcal{K}}\alpha_k R_k$.
At each iteration, Step 4 updates the auxiliary variables $\{\mathbf{u},\mathbf{w},\boldsymbol{\Sigma}\}$ using (\ref{eq:optimal_sigma}) and (\ref{eq:optimal_u_w}) to transform the original weighted sum-rate maximization task \eqref{prob:original_fullyd} into the equivalent WMMSE formalism \eqref{prob:WMMSE_fullyd}. In Step 5, we initialize the vector $\mathbf{c}$ as $\mathbf{c} = \sum_{m\in\mathcal{M}} (\mathbf{I}_K\otimes\mathbf{U} \widehat{\mathbf{H}}_m^H) \mathbf{v}_{A,m}$. This allows us to reduce the computational complexity for calculating $\mathbf{b}_{i}$ in \eqref{eq:bi} as described shortly.

The dual problem in \eqref{prob:RWMMSE_dual_problem2} is addressed for each AP $i\in\mathcal{M}$ based on the ellipsoid method in Steps 9-15. To this end, Step 7 initializes the ellipsoid and the best dual variables $\boldsymbol{\rho}_{i}^{\text{best}}$, which is followed by the update of the vector $\tilde{\mathbf{c}}$ in Step 8. As a consequence, $\mathbf{b}_{i}$ in \eqref{eq:bi} can be sequentially updated as
\begin{align} \label{eq:b_iwithc}
    \mathbf{b}_i = -\mathbf{W}\widehat{\mathbf{h}}_{A,i} + (\mathbf{I}_K \otimes \widehat{\mathbf{H}}_i) \tilde{\mathbf{c}}.
\end{align}
This \textit{incremental update} strategy  based on $\mathbf{c}$ and $\tilde{\mathbf{c}}$ allows the complexity of computing $\,$ $\sum_{m\in\mathcal{M}\setminus\{i\}} ( \mathbf{I}_K\otimes \mathbf{U}\widehat{\mathbf{H}}_m^H ) \mathbf{v}_{A,m}$ for $\mathbf{b}_i$ in \eqref{eq:bi} over all $i\in\mathcal{M}$ to scale linearly with $M$, rather than $M^2$.

At each iteration of the ellipsoid method, Step 11 obtains the primal variables $\{\mathbf{v}_{A,i},\boldsymbol{\Omega}_{i}\}$ using \eqref{eq:v_k_decoupled} and \eqref{eq:optimal_solution_quantization} with fixed dual variables $\boldsymbol{\rho}_{i}=\boldsymbol{\rho}_{i}^{(t-1)}$. According to the feasibility of $\boldsymbol{\rho}_{i}^{(t-1)}$, Step 12 computes the subgradient vector $\mathbf{m}_{i}^{(t-1)}$, which is followed by the ellipsoid updates in Step 13. Also, in Step 14, we keep the best dual variables $\boldsymbol{\rho}_{i}^{\text{best}}$ that achieve the largest dual function.
Such ellipsoid update procedures are repeated until the determinant of the ellipsoid matrix $\boldsymbol{\Psi}_i$ falls below the threshold $\epsilon_{\mathrm{in}}$.

Once the ellipsoid method is completed, Step 16 refines the primal variables $\{\mathbf{v}_{A,i}, \boldsymbol{\Omega}_i\}$ with the best dual variables $\boldsymbol{\rho}_{i}^{\text{best}}$. Accordingly, the vector $\mathbf{c}$ is modified using the updated beamforming vector $\mathbf{v}_{A,i}$. This is leveraged to solve the per-AP optimization problem \eqref{prob:RWMMSE_fullyd} for the subsequent APs. 
After collecting the per-AP solutions, we compute the updated weighted sum-rate $R_{\mathrm{sum}}^{\mathrm{new}}$. This sequential optimization process is repeated until the weighted sum-rate variation $R_{\mathrm{var}} = |R_{\mathrm{sum}}^{\mathrm{new}}-R_{\mathrm{sum}}^{\mathrm{old}}|$ falls below the threshold $\epsilon_{\mathrm{out}}$. If $R_{\mathrm{var}}>\epsilon_{\mathrm{out}}$, we update $R_{\mathrm{sum}}^{\mathrm{old}}\leftarrow R_{\mathrm{sum}}^{\mathrm{new}}$ and proceed to the next iteration.

\subsection{Convergence Behavior} \label{sub:convergence}

For fixed primal variables $\{\mathbf{v},\boldsymbol\Omega\}$, the closed-form updates of the auxiliary variables $\{\mathbf{u},\mathbf{w},\boldsymbol\Sigma\}$ in \eqref{eq:optimal_u_w_sigma} optimally solve their corresponding blocks in the equivalent WMMSE problem and therefore do not increase the WMMSE objective. Conversely, for fixed auxiliary variables, the primal-variable block is optimized through a sequence of equivalent per-AP subproblems, where $\{\mathbf{v}_{A,i},\boldsymbol\Omega_i\}$ are updated while the other AP variables are fixed. 
Each per-AP subproblem is convex and satisfies Slater's condition, and thus strong duality holds. 
Let $\mathcal{L}_i(\mathbf{p}_i,\boldsymbol\rho_i)$ denote the corresponding Lagrangian, where $\mathbf{p}_i=\{\mathbf{v}_{A,i},\boldsymbol\Omega_i\}$ and $\boldsymbol\rho_i=\{\lambda_i,\mu_i\}$. Under the assumption that each per-AP subproblem is solved optimally, the optimal primal-dual pair $(\mathbf{p}_i^\star,\boldsymbol\rho_i^\star)$ satisfies
\begin{align}
\mathcal{L}_i(\mathbf{p}_i^\star,\boldsymbol\rho_i)
\leq
\mathcal{L}_i(\mathbf{p}_i^\star,\boldsymbol\rho_i^\star)
\leq
\mathcal{L}_i(\mathbf{p}_i,\boldsymbol\rho_i^\star), \label{eq:saddle-point-optimality}
\end{align}
which establishes the saddle-point optimality of each exact per-AP update. The dual variables are computed using the ellipsoid method, whose search region is successively reduced using valid subgradients \cite{Nedic:JOTA09}.
Since both the optimal auxiliary-variable updates and the exact per-AP updates do not increase the WMMSE objective, the resulting objective sequence is monotonically non-increasing and bounded below, and therefore converges. Under the standard regularity conditions required for block coordinate descent, any limit point of the variable sequence generated by these exact updates satisfies the first-order stationarity conditions of the equivalent WMMSE problem.

In practice, the ellipsoid iterations are stopped when the numerical stopping criterion is met, yielding approximate per-AP solutions. The theoretical guarantees above assume exact per-AP solutions and do not, in general, establish exact monotonicity or convergence to a stationary point for an implementation using a fixed nonzero stopping tolerance.

\subsection{Complexity Discussion} \label{sub:complexity}

In the conventional WMMSE algorithm \cite{Yu:WCL19}, Steps 5--15 of Algorithm 1, which optimizes $\{\mathbf{v}, \boldsymbol{\Omega}\}$, are executed without the per-AP decomposition presented in Sec. \ref{sub:per_AP_problem}. Thus, the complexity of updating $\{\mathbf{v}, \boldsymbol{\Omega}\}$ is determined by a large-scale high-dimensional convex program and is upper bounded by $\mathcal{O}\left( n_V \left( n_V^3 + n_O\right) \right)$, where $n_V = \mathcal{O}(M N_A \left( N_A + K \right))$ denotes the number of optimization variables, and $n_O = \mathcal{O}(M N_A K \left( M^2 N_A^2 + K\right))$ represents the number of arithmetic operations for computation of the objective function and constraints~\cite[p. 4]{Ben-Tal}. Therefore, the complexity of the conventional WMMSE increases proportionally to $M^{4}$, thereby limiting its implementation in practical CF-mMIMO systems with a large number of APs.

In contrast, the proposed R-WMMSE algorithm decomposes the joint optimization of $\{\mathbf{v},\boldsymbol\Omega\}$ into $M$ per-AP subproblems.
The complexity of updating the auxiliary variables $\{\mathbf{u},\mathbf{w},\boldsymbol\Sigma\}$ in Step 4 scales as $\mathcal{O}(MK^2N_A^2+MN_A^3)$, where the first term mainly arises from evaluating the interference and CSI-error quadratic terms, while the second term accounts for the matrix factorizations associated with $\boldsymbol\Sigma$. The aggregate vector $\mathbf{c}$ in Step 5 can be computed with complexity $\mathcal{O}(MK^2N_A)$ by exploiting its block-diagonal structure.
For the primary-variable update for each AP, the dominant operation in each ellipsoid iteration is the update of $\mathbf{v}_{A,i}$ in Step 11 using \eqref{eq:v_k_decoupled}, which incurs $\mathcal{O}(N_A^3+KN_{A}^{2})$ operations. 
Computing the dual subgradient in Step 12 requires $\mathcal{O}(N_A^3+KN_A^2)$ operations, whereas the update of the two-dimensional ellipsoid in Step 13 has constant complexity. Moreover, the aggregate-vector subtraction and refresh operations in Steps 8 and 17 each require $\mathcal{O}(K^2N_A)$ operations and prevent an additional factor of $M$ that would otherwise arise from explicitly summing the contributions from all other APs.
Combining these operations, the overall complexity of Algorithm 1 is given by $\mathcal{O}\left(T_{\mathrm{O}}M[K^2N_A^2+
T_{\mathrm{E}}(N_A^3+KN_{A}^{2})]\right)$, where $T_{\mathrm{E}}$ and $T_{\mathrm{O}}$ denote the numbers of per-AP ellipsoid iterations and outer R-WMMSE iterations, respectively.
The convergence behavior and substantial complexity savings achieved by the proposed R-WMMSE algorithm will be numerically demonstrated in Sec. \ref{sec:results}.

\subsection{Signaling Overhead and Fronthaul Latency} \label{sub:fronthaul-overhead}

In this subsection, we discuss the signaling overhead and the corresponding fronthaul latency during the channel training phase, in which each AP $i$ sends either the received pilot sequences or the estimated channel vectors to the CP over the fronthaul link. To this end, we consider two different strategies \cite{Kang:TWC14}: \textit{forward-then-estimate} (FE) and \textit{estimate-then-forward} (EF).

\subsubsection{FE Strategy} \label{subsub:forward-then-estimate}

In the FE strategy, each AP $i$ forwards the received pilot sequences to the CP, where the channel vectors are estimated. 
Since the received pilot signals consist of $L N_A$ complex scalars, the amount of information to be transferred over each fronthaul link is $2 L N_A N_q$ bits when each in-phase or quadrature component is quantized using $N_q$ bits. Consequently, the corresponding fronthaul latency is given by
\begin{align}
    T_{\text{FE},i} = \frac{2 L N_A N_q}{C_i} \,\, \text{sec}. 
\end{align}
%$T_{\text{FE}} = 2 L N_A N_q / C_F$ sec.}

\subsubsection{EF Strategy} \label{subsub:estimate-then-forward}

In the EF strategy, each AP $i$ first estimates its local CSI and forwards the estimated channel vectors $\{\widehat{\mathbf{h}}_{k,i}\}_{k\in\mathcal{K}}$ to the CP. Since the local CSI consists of $K N_A$ complex scalars, the amount of information to be transferred over each fronthaul link is $2 K N_A N_q$ bits, resulting in a fronthaul latency of
\begin{align}
    T_{\text{EF},i} = \frac{2 K N_A N_q}{C_i} \,\, \text{sec}.
\end{align}
Comparing the two strategies, the FE strategy incurs a lower fronthaul latency than EF when the pilot sequence length is smaller than the number of UEs (i.e., $L< K$). Otherwise, the EF strategy is more efficient.

\section{Extensions to Hybrid Analog-Digital Architecture} \label{sec:hybrid_extension}

So far, we have focused on optimizing the digital beamforming vectors $\mathbf{v}$. Such a fully-digital beamforming system assumes that the number of RF chains at each AP, denoted by $N_{R}$, equals that of antennas $N_{A}$. In practice, however, we have $N_{R}<N_{A}$ due to the high cost associated with increasing the number of RF chains~\cite{Molisch:COMMA17, Sohrabi:JSTSP16, Liang:WCL14, Kim:TCOM19}. Therefore, it is necessary to develop hybrid analog-digital beamforming strategies at APs.

As shown in (\ref{eq:fully_beamforming}), in the CF-mMIMO system, the CP sends the digital-beamformed and fronthaul-compressed signal vector of length $N_{R}$ to the APs. Then, each AP $i$ employs the analog beamformer $\mathbf{A}_i\in\mathbb{C}^{N_A\times N_R}$. For clarity, in the hybrid beamforming system, the digital beamforming vector and the quantization noise of AP $i$ are respectively denoted by $\mathbf{v}_{k,i}^{\mathrm{hyb}} \in\mathbb{C}^{N_R\times 1}$ and $\mathbf{q}_{i}^{\mathrm{hyb}}\sim\mathcal{CN}(\mathbf{0},\boldsymbol{\Omega}_i^{\mathrm{hyb}})\in\mathbb{C}^{N_R\times 1}$, where $\boldsymbol{\Omega}_i^{\mathrm{hyb}}\in\mathbb{C}^{N_R\times N_R}$ accounts for the quantization noise covariance matrix.
Accordingly, the transmitted signal vector at AP $i$ is refined as 
\begin{align}
    \mathbf{x}_i = \sum_{k \in \mathcal{K}} \mathbf{A}_i \mathbf{v}_{k,i}^{\mathrm{hyb}} s_k + \mathbf{A}_i \mathbf{q}_i^{\mathrm{hyb}}.
    \label{eq:hybrid_beamforming_after}
\end{align}
The analog beamformer $\mathbf{A}_i$ is implemented using phase shifters~\cite{An:ICC23}, and thus its elements are subject to constant modulus constraints as
\begin{align}
    \big|[\mathbf{A}_i]_{n,n'}\big| = 1, \:\forall (i,n,n') \in \mathcal{M} \times \mathcal{N} \times \mathcal{R},
        \label{eq:hybrid_modulus_constraints}
\end{align}
where $\mathcal{N}=\{1, \dots, N_A\}$ and $\mathcal{R}=\{1, \dots, N_R\}$ denote the index sets of antennas and RF chains, respectively.

With the above hybrid beamforming model, the achievable data rate for UE $k$ is given by
\begin{align}
    R_k^{\mathrm{hyb}} = \log_2\left(1 + \frac{|\widehat{\mathbf{h}}_k^H \bar{\mathbf{A}} \mathbf{v}_k^{\mathrm{hyb}}|^2}{\text{IF}_k^{\mathrm{hyb}}(\mathbf{v}^{\mathrm{hyb}}, \boldsymbol{\Omega}^{\mathrm{hyb}}, \mathbf{A})} \right),
    \label{eq:rate_hybrid}
\end{align}
where $\text{IF}_k^{\mathrm{hyb}}(\mathbf{v}^{\mathrm{hyb}}, \boldsymbol{\Omega}^{\mathrm{hyb}}, \mathbf{A})$ is defined as
\begin{align}
    \text{IF}_k^{\mathrm{hyb}}&(\mathbf{v}^{\mathrm{hyb}}, \boldsymbol{\Omega}^{\mathrm{hyb}}, \mathbf{A}) = N_0^{\text{dl}} + \sum_{l \in \mathcal{K}\setminus \{k\}} \left|\widehat{\mathbf{h}}_k^H \bar{\mathbf{A}} \mathbf{v}_l^{\mathrm{hyb}}\right|^2 \nonumber \\
    & + \sum_{l \in \mathcal{K}} (\bar{\mathbf{A}} \mathbf{v}_l^{\mathrm{hyb}})^H \mathbf{E}_k \bar{\mathbf{A}} \mathbf{v}_l^{\mathrm{hyb}} + \widehat{\mathbf{h}}_k^H \bar{\mathbf{A}} \bar{\boldsymbol{\Omega}}^{\mathrm{hyb}} \bar{\mathbf{A}}^H \widehat{\mathbf{h}}_k \nonumber \\
    & + \operatorname{tr}\left(\bar{\mathbf{A}}^H \mathbf{E}_k \bar{\mathbf{A}} \bar{\boldsymbol{\Omega}}^{\mathrm{hyb}}\right) 
    \label{eq:IF_func_hybrid}
\end{align}
with $\bar{\boldsymbol{\Omega}}^{\mathrm{hyb}}\!\!=\!\!\text{blkdiag}(\boldsymbol{\Omega}_1^{\mathrm{hyb}}, \dots, \boldsymbol{\Omega}_M^{\mathrm{hyb}})$, $\mathbf{v}^{\mathrm{hyb}}\!\!=\!\!\{\mathbf{v}_{k,i}^{\mathrm{hyb}}\}_{k\in\mathcal{K}, i\in\mathcal{M}}$, $\boldsymbol{\Omega}^{\mathrm{hyb}}=\{\boldsymbol{\Omega}_i^{\mathrm{hyb}}\}_{i\in\mathcal{M}}$, $\mathbf{A}=\{\mathbf{A}_i\}_{i\in\mathcal{M}}$, and $\bar{\mathbf{A}} = \text{blkdiag}(\mathbf{A}_1, \dots, \mathbf{A}_M)$.

Therefore, the joint optimization problem for the hybrid analog-digital beamforming and fronthaul compression is formulated as 
\begin{subequations} \label{prob:hybrid}
\begin{align}
    \underset{\mathbf{v}^{\mathrm{hyb}}, \boldsymbol{\Omega}^{\mathrm{hyb}}, \mathbf{A}}{\max}&\:\: \sum_{k \in \mathcal{K}} \alpha_k R_k^{\mathrm{hyb}} \label{prob:obj_func_hybrid} \\
    \text{s.t.}&\:\: \mathrm{(\ref{prob:fc_fullyd}), (\ref{eq:hybrid_modulus_constraints})}, \label{prob:modulus_fronthaul_hybrid} \\
    &\:\: \sum_{k \in \mathcal{K}} \operatorname{tr}(\mathbf{A}_i \mathbf{v}_{k,i}^{\mathrm{hyb}} (\mathbf{v}_{k,i}^{\mathrm{hyb}})^H \mathbf{A}_i^H) \nonumber \\
    &\quad + \operatorname{tr}(\mathbf{A}_i \boldsymbol{\Omega}_i^{\mathrm{hyb}} \mathbf{A}_i^H) \leq P_i, \:\forall i \in \mathcal{M},
        \label{prob:hybrid_power_constraints}
\end{align}    
\end{subequations}
where \eqref{prob:hybrid_power_constraints} indicates the per-AP power constraint under the hybrid beamforming architecture.

Compared to the fully-digital beamforming scenario studied in Sec. \ref{sec:proposed_fullyd}, problem (\ref{prob:hybrid}) presents additional challenges due to the multiplicative coupling between the analog beamformer $\mathbf{A}$ and the digital processing variables $\{\mathbf{v}^{\mathrm{hyb}}, \boldsymbol{\Omega}^{\mathrm{hyb}}\}$, as well as the presence of the non-convex constant modulus constraints (\ref{eq:hybrid_modulus_constraints}).
To address these challenges, we adopt the WMMSE-based reformulation by introducing auxiliary variables $\{\mathbf{u}, \mathbf{w}, \boldsymbol{\Sigma}\}$. Then, the analog and digital processing variables are iteratively optimized based on the AO procedure. Notice that for fixed analog beamformer $\mathbf{A}$, the digital variables $\{\mathbf{v}_{A,i}^{\mathrm{hyb}}, \boldsymbol{\Omega}_{i}^{\mathrm{hyb}}\}_{i\in\mathcal{M}}$ can be determined using the proposed R-WMMSE method in Algorithm \ref{algorithm1}. Here, we modify the update rules in \eqref{eq:v_k_decoupled} and \eqref{eq:optimal_solution_quantization} as
\begin{subequations}
\begin{align}
    &\mathbf{v}_{k,i}^{\mathrm{hyb}} = -\mathbf{T}_{i}^{\mathrm{hyb}}(\boldsymbol{\rho}_{i})^{-1} \mathbf{b}_{k,i}^{\mathrm{hyb}},\\
    &\boldsymbol{\Omega}_i^{\mathrm{hyb}} = \mu_i \mathbf{T}_{i}^{\mathrm{hyb}}(\boldsymbol{\rho}_{i})^{-1},
\end{align}
\end{subequations}
where $\mathbf{T}_{i}^{\mathrm{hyb}}(\boldsymbol{\rho}_{i})\!=\!\mathbf{A}_i^H \mathbf{D}_i \mathbf{A}_i+ \lambda_i \mathbf{A}_i^H \mathbf{A}_i + \mu_i \boldsymbol{\Sigma}_i^{-1}$
, $\mathbf{b}_i^{\mathrm{hyb}}\!=\!-(\mathbf{I}_K \otimes \mathbf{A}_i^H )\mathbf{W}\widehat{\mathbf{h}}_{A,i} + (\mathbf{I}_K \otimes \mathbf{A}_i^H \widehat{\mathbf{H}}_i)\tilde{\mathbf{c}}^{\mathrm{hyb}}$,  and $\tilde{\mathbf{c}}^{\mathrm{hyb}} \!\!=\!\! \sum_{m\in\mathcal{M}\setminus\{i\}}(\mathbf{I}_K\otimes\mathbf{U} \widehat{\mathbf{H}}_m^H \mathbf{A}_{m}) \mathbf{v}_{A,m}^{\mathrm{hyb}}$. Here, $\mathbf{b}_{k,i}^{\mathrm{hyb}}\in\mathbb{C}^{N_{R}\times1}$ denotes the $k$th subvector of $\mathbf{b}_{i}^{\mathrm{hyb}}$ and $\tilde{\mathbf{c}}^{\mathrm{hyb}}$ can be computed using the same subtract--update--add strategy as in the fully-digital case.

We focus on the optimization of $\mathbf{A}$ while keeping the digital processing variables $\{\mathbf{v}^{\mathrm{hyb}}, \boldsymbol{\Omega}^{\mathrm{hyb}}\}$ fixed.
To enforce the constant modulus constraints (\ref{eq:hybrid_modulus_constraints}), 
we parameterize each analog beamforming element $[\mathbf{A}_i]_{n,n'}$ by its phase $\theta_{i,n,n'}$ as
\begin{align}
    [\mathbf{A}_i]_{n,n'} = \exp{\left(j\theta_{i,n,n'}\right)}, \: \forall (n,n') \in \mathcal{N}\times\mathcal{R}.
    \label{eq:parameterize_theta}
\end{align}
For notational convenience, we define the phase shift matrix $\boldsymbol{\Theta}_i \in \mathbb{R}^{N_A \times N_R}$ with $[\boldsymbol{\Theta}_i]_{n,n'}=\theta_{i,n,n'}$ and the set of all phase shift matrices $\boldsymbol{\Theta}=\{ \boldsymbol{\Theta}_i\}_{i\in\mathcal{M}}$. Then, under the WMMSE framework, the problem of updating the phase shift matrix $\boldsymbol{\Theta}$ for fixed $\{\mathbf{v}^{\mathrm{hyb}}, \boldsymbol{\Omega}^{\mathrm{hyb}}, \mathbf{u}, \mathbf{w}, \boldsymbol{\Sigma}\}$ is stated as
\begin{subequations} \label{prob:WMMSE_hybrid}
    \begin{align}
    \underset{\boldsymbol{\Theta}}{\min} & \:\: f(\boldsymbol{\Theta})=\sum_{k \in \mathcal{K}} \alpha_k w_k \text{MSE}_k^{\mathrm{hyb}}(\mathbf{v}^{\mathrm{hyb}}, \boldsymbol{\Omega}^{\mathrm{hyb}},\mathbf{A},u_k) \label{prob:WMMSE_objective_hybrid_analog} \\
    \mathrm{s.t.} &\:\: \mathrm{(\ref{prob:hybrid_power_constraints}),(\ref{eq:parameterize_theta})},
\end{align}    
\end{subequations}
where $\text{MSE}_k^{\mathrm{hyb}}(\mathbf{v}^{\mathrm{hyb}},\boldsymbol{\Omega}^{\mathrm{hyb}},\mathbf{A},u_k) = |1-u_k^* \widehat{\mathbf{h}}_k^H \bar{\mathbf{A}}\mathbf{v}_k^{\mathrm{hyb}}|^2+|u_k|^2\, \text{IF}_k^{\mathrm{hyb}}(\mathbf{v}^{\mathrm{hyb}}, \boldsymbol{\Omega}^{\mathrm{hyb}},\mathbf{A})$.
In (\ref{prob:WMMSE_hybrid}), the fronthaul capacity constraints (\ref{prob:fc_fullyd}) are excluded, since they are not affected by the choice of the analog beamformers $\mathbf{A}$. Substituting the parameterization (\ref{eq:parameterize_theta}) into (\ref{prob:WMMSE_hybrid}) removes the constant modulus constraints (\ref{eq:hybrid_modulus_constraints}).

Gradient-based approaches~\cite{Chang:TWC24,Yu:JSTSP16} are commonly used to update the analog beamformer in hybrid beamforming design. However, directly applying such methods requires substantial computational overhead for evaluating gradients from globally coupled expressions, especially as the number of APs increases. To address this issue, we apply the proposed R-WMMSE framework to the analog phase update, enabling the analog beamformer to be updated within a per-AP subproblem while reusing precomputed quantities. Specifically, by leveraging the per-AP decomposition, the subproblem associated with the analog beamformer $\mathbf{A}_i$ at AP $i$ can be written as 
\begin{subequations} \label{prob:RWMMSE_hybrid_analog}
    \begin{align}
    \underset{\boldsymbol{\Theta}_i}{\min} & \:\: \operatorname{tr}\left( \mathbf{A}_i^H \mathbf{D}_i \mathbf{A}_i (\mathbf{V}_i +\boldsymbol{\Omega}_i^{\mathrm{hyb}})\right) + 2 \Re \big\{ \operatorname{tr}( \mathbf{A}_i^H \mathbf{F}_i )\big\} \label{prob:RWMMSE_objective_hybrid_analog} \\
    \mathrm{s.t.} &\:\: \operatorname{tr}\left( \mathbf{A}_i^H \mathbf{A}_i (\mathbf{V}_i + \boldsymbol{\Omega}_i^{\mathrm{hyb}}) \right) \leq P_i,\label{prob:RWMMSE_powerC_hybrid_analog}
\end{align}    
\end{subequations}
where
\begin{subequations}
    \begin{align}
        \!\!\!\mathbf{V}_i &=\sum_{k\in\mathcal{K}} \mathbf{v}_{k,i}^{\mathrm{hyb}}(\mathbf{v}_{k,i}^{\mathrm{hyb}})^H, \\
        \!\!\!\mathbf{F}_i &= \sum_{k\in\mathcal{K}} \left( -\alpha_k w_k u_k \widehat{\mathbf{h}}_{k,i} (\mathbf{v}_{k,i}^{\mathrm{hyb}})^H \!\! + \!\widehat{\mathbf{H}}_i \mathbf{U} \tilde{\mathbf{f}}_{k,i} (\mathbf{v}_{k,i}^{\mathrm{hyb}})^H \right), \\
        \!\!\!\tilde{\mathbf{f}}_{k,i} &= \sum_{m\in\mathcal{M}\setminus\{i\}} \widehat{\mathbf{H}}_m^H \mathbf{A}_m \mathbf{v}_{k,m}^{\mathrm{hyb}}.
    \end{align}
\end{subequations}
To further reduce the computational overhead in evaluating $\tilde{\mathbf{f}}_{k,i}$, we adopt the same aggregation strategy used for the digital beamforming update. Specifically, we define the global aggregate $\mathbf{f}_k=\sum_{m\in\mathcal{M}}\widehat{\mathbf{H}}_m^H \mathbf{A}_m \mathbf{v}_{k,m}^{\mathrm{hyb}}, \forall k \in \mathcal{K}$, so that the term excluding AP $i$ can be obtained as $\tilde{\mathbf{f}}_{k,i}=\mathbf{f}_k-\widehat{\mathbf{H}}_i^H \mathbf{A}_i \mathbf{v}_{k,i}^{\mathrm{hyb}}, \forall k \in \mathcal{K}$. During the per-AP update for AP $i$, $\mathbf{f}_k$ is cached, and $\tilde{\mathbf{f}}_{k,i}$ is obtained via the above subtraction, rather than recomputing the summation over all APs. After updating $\mathbf{A}_i$~(equivalently, $\boldsymbol{\Theta}_i$), the aggregate is refreshed by adding back the updated contribution, i.e., $\mathbf{f}_k\leftarrow\tilde{\mathbf{f}}_{k,i} + \widehat{\mathbf{H}}_i^H \mathbf{A}_i \mathbf{v}_{k,i}^{\mathrm{hyb}}, \forall k \in \mathcal{K}$.

Let $f_{\mathbf{A}_i}(\boldsymbol{\Theta}_i)$ denote the objective function for updating the analog beamformer $\mathbf{A}_i$ at AP $i$, defined as $f_{\mathbf{A}_i}(\boldsymbol{\Theta}_i) = \operatorname{tr}\left( \mathbf{A}_i^H \mathbf{D}_i \mathbf{A}_i (\mathbf{V}_i +\boldsymbol{\Omega}_i^{\mathrm{hyb}})\right) + 2 \Re \{ \operatorname{tr}( \mathbf{A}_i^H \mathbf{F}_i )\}$.
The partial derivative with respect to each phase element $\theta_{i,n,n'}$ is given by
\begin{align}\label{eq:derivative_theta}
    \frac{\partial f_{\mathbf{A}_i}(\boldsymbol{\Theta}_i)}{\partial \theta_{i,n,n'}} = 2 \Re \Big\{ j \exp{ (j \theta_{i,n,n'})} \big[ \boldsymbol{\Phi}_i \big]_{n,n'}  \Big\},
\end{align}
where
\begin{align}\label{eq:derivative_theta_Phi}
    \boldsymbol{\Phi}_i =  \left(\mathbf{D}_i \mathbf{A}_i (\mathbf{V}_i + \boldsymbol{\Omega}_i^{\mathrm{hyb}}) + \mathbf{F}_i\right)^H.
\end{align}
With the above partial derivative, we update each phase element via a gradient descent method as
\begin{align}
    \theta_{i,n,n'} \leftarrow \theta_{i,n,n'} - \tau  \frac{\partial  f_{\mathbf{A}_i}(\boldsymbol{\Theta}_i)}{\partial \theta_{i,n,n'}}. \label{eq:update_theta}
\end{align}
Here, $\tau \in (0,1)$ is a diminishing step size updated as $\tau \leftarrow \tau r$ with a prescribed rate $r\in(0,1)$. This geometric decay yields progressively smaller updates and promotes stable convergence of the analog phase refinement.
The phase gradient descent iterations continue until the absolute difference between consecutive values of each phase shift falls below the threshold $\epsilon_{\mathrm{GD}}$.
After completing all phase-shift updates of $\mathbf{A}_i$, the per-AP power constraint is enforced by projection. Specifically, if
$\operatorname{tr}\left(\mathbf{A}_i^H\mathbf{A}_i
\left(\mathbf{V}_i+\boldsymbol{\Omega}_i^{\mathrm{hyb}}\right)\right)>P_i$,
we update
\begin{subequations} \label{eq:projection_power_hyb}
    \begin{align} 
        \mathbf{v}_{k,i}^{\mathrm{hyb}} &\leftarrow \sqrt{\xi_i}\mathbf{v}_{k,i}^{\mathrm{hyb}}, \: \forall k \in \mathcal{K}, \\
        \boldsymbol{\Omega}_i^{\mathrm{hyb}} &\leftarrow
        \xi_i\boldsymbol{\Omega}_i^{\mathrm{hyb}}, 
    \end{align}    
\end{subequations}
where
$\xi_i= P_i / \operatorname{tr}(\mathbf{A}_i^H\mathbf{A}_i
(\mathbf{V}_i+\boldsymbol{\Omega}_i^{\mathrm{hyb}}))$.

Next, we discuss the computational complexity of the proposed R-WMMSE algorithm in the hybrid beamforming architecture. For fixed analog beamformers $\mathbf{A}$, the overall complexity of updating the digital processing variables $\{\mathbf{v}^{\mathrm{hyb}}, \boldsymbol{\Omega}^{\mathrm{hyb}}\}$ is given by
$\mathcal{O}(T_{\mathrm{O}}M[N_A^2N_R+KN_A^2+K^2N_A$ $+T_{\mathrm{E}}(N_R^3+KN_R^2)])$.
Compared with the corresponding ellipsoid-dependent complexity $\mathcal{O}(N_A^3+KN_A^2)$ in the fully-digital case, this provides a substantial complexity reduction since $N_R\ll N_A$ in hybrid systems.
Additionally, the complexity of updating the analog beamformers $\mathbf{A}$ for fixed $\{\mathbf{v}^{\mathrm{hyb}},\boldsymbol{\Omega}^{\mathrm{hyb}}\}$ is given by $\mathcal{O}\left(T_{\mathrm{G}}M(N_A^2N_R+KN_A^2+K^2N_A)\right)$, where $T_{\mathrm{G}}$ denotes the number of gradient-descent iterations for the analog beamformer update. Specifically, by exploiting the aggregated vector $\mathbf{f}_k$ within the proposed R-WMMSE approach, the complexity of the analog beamformer update scales linearly with the number of APs $M$. It should also be noted that the phase elements associated with each AP can be updated in parallel. In contrast, a conventional gradient-based analog beamformer update, e.g., \cite{An:ICC23}, typically incurs a significantly higher computational cost, scaling as $\mathcal{O}\left(T_{\mathrm{G}}M^2N_A^2N_R^2K(K+N_R)\right)$. Therefore, the proposed R-WMMSE-based analog beamformer update provides a clear computational advantage in terms of asymptotic complexity.

\section{Numerical Results} \label{sec:results}

We present numerical results to validate the effectiveness of the proposed R-WMMSE algorithm.
In our setup, UEs and APs are uniformly distributed within a circular region of radius 350 m. The large-scale path-loss between UE $k$ and AP $i$ is modeled as
$\rho_{k,i} = (d_{k,i}/d_0)^{-3}$,
where $d_0 = 30$ m is the reference distance and $d_{k,i}$ denotes the distance between UE $k$ and AP $i$.
To focus on analyzing the impact of downlink SNR, fronthaul capacity, and network size, we assume an uncorrelated fading model, i.e., $\mathbf{R}_{k,i} = \rho_{k,i}\mathbf{I}$.
% \textcolor{teal}{(Comment 1.2a?)}
Unless stated otherwise, the transmit power and fronthaul capacity of all APs are assumed to be identical, i.e., $P_i = P$, $\forall i\in\mathcal{M}$ and $C_i = C$, $\forall i\in\mathcal{M}$.
The pilot sequences are assigned to the UEs in a round-robin manner, where the pilot index of UE $k$ is given by 
$l_k = \left((k-1) \mod L\right) + 1$ for $k\in\mathcal{K}$.
The downlink and uplink signal-to-noise ratios (SNRs) are defined respectively as $\text{SNR}^{\text{dl}}\triangleq P/N_0^{\text{dl}}$ and $\text{SNR}^{\text{ul}}\triangleq P^{\text{ul}}/N_0^{\text{ul}}$. 
To evaluate the weighted sum-rate objective, the user weights are randomly generated and then normalized to keep a consistent scale across realizations. Specifically, we draw $\tilde{\alpha}_k\sim\mathcal{U}(0,1)$ independently for each $k\in\mathcal{K}$ and set
\begin{align}
    \alpha_k = \frac{K \tilde{\alpha}_k}{\sum_{j\in\mathcal{K}}\tilde{\alpha}_j}, \:\: \forall k \in \mathcal{K}.
\end{align}
The use of randomly generated weights allows us to evaluate the proposed R-WMMSE framework with arbitrary nonnegative UE weights. More generally, the framework can readily accommodate practical weighting strategies, such as proportional-fair (PF) scheduling \cite{Viswanath:TIT02}.
Throughout the simulations, the proposed algorithms are initialized with randomly selected initial points. Although the converged performance can depend on the initialization due to the non-convexity of the considered optimization problems, we observed numerically that the performance variations across different initializations are minor for the relatively large network with $M=32$ APs, which is predominantly considered in this section. Therefore, we use a single randomly selected initial point for the simulations throughout this section.
The weighted sum-rates are reported in bps by multiplying the SE in (\ref{eq:rate_fullyd}) by the system bandwidth set to 20 MHz.
Each data point is averaged over 100 independent channel realizations. The stopping thresholds are set to $\epsilon_{\mathrm{out}}=10^{-4}$, $\epsilon_{\mathrm{in}}=10^{-20}$, and $\epsilon_{\mathrm{GD}}=10^{-8}$.
All simulations are conducted in MATLAB on a workstation equipped with an Intel Core i9-10900K CPU (3.70 GHz) and 128 GB of RAM.
%, running a 64-bit operating system.
%The reported runtimes correspond to the MATLAB implementation executed on the CPU.

\subsection{Fully-Digital Scenario ($N_R = N_A$)} \label{sub:numerical-fully-digital}

In this subsection, we consider fully-digital scenarios, where each AP has $N_R=N_A$ RF chains and does not need to employ analog beamforming.
For comparison, we consider the following benchmarks.
\begin{itemize}
    \item \textit{Conventional WMMSE}: The conventional WMMSE method \cite{Yu:WCL19, Christensen:TWC08} described in Sec. \ref{sec:problem_formulation} serves as a performance benchmark for the proposed algorithm. 
    Specifically, Steps 5--18 of Algorithm 1 are replaced with an update of $\{\mathbf{v}, \boldsymbol{\Omega}\}$ obtained by solving problem (\ref{prob:WMMSE_fullyd}) using CVX \cite{Grant:CVX20}, with $\{\mathbf{u}, \mathbf{w}, \boldsymbol{\Sigma}\}$ fixed.
    \item \textit{Maximum-ratio transmission (MRT)}: The beamforming vector is set to $\mathbf{v}_{k,i}=p_{k,i}\widehat{\mathbf{h}}_{k,i}/\|\widehat{\mathbf{h}}_{k,i}\|$, where $p_{k,i}$ with $\sum_{k\in\mathcal{K}}p_{k,i}\leq P$ accounts for the transmit power of AP $i$ assigned to UE $k$. The noise covariance along with the power allocation variables $\{p_{k,i}\}_{k\in\mathcal{K},i\in\mathcal{M}}$ are jointly optimized using the conventional WMMSE approach \cite{Lo:ICC99}.
    \item \textit{Zero-forcing (ZF)}: The ZF beamforming is adopted as $\mathbf{v}_{k,i}=p_{k,i}\tilde{\mathbf{h}}_{k,i}/\|\tilde{\mathbf{h}}_{k,i}\|$, where $\tilde{\mathbf{h}}_{k,i}$ stands for the $k$th column of $\widehat{\mathbf{H}}_{i}^{H}(\widehat{\mathbf{H}}_{i}\widehat{\mathbf{H}}_{i}^{H})^{-1}$. Similar to the MRT baseline, we optimize the power control and noise covariances using the conventional WMMSE algorithm \cite{Spencer:TSP04}.
    \item \textit{Local signal-to-leakage-plus-noise ratio (L-SLNR)}: The L-SLNR beamforming is adopted as $\mathbf{v}_{k,i}=p_{k,i}\mathbf{w}_{k,i}/\|\mathbf{w}_{k,i}\|$, where $\mathbf{w}_{k,i}$ is given as $\mathbf{w}_{k,i} = (\sum_{k^{\prime}\in\mathcal{K}\setminus\{k\}} \widehat{\mathbf{h}}_{k^{\prime},i}\widehat{\mathbf{h}}_{k^{\prime},i}^H + \mathbf{E}_{k,i} + N_0^{\text{dl}} \mathbf{I} )^{-1}  \widehat{\mathbf{h}}_{k,i}$, and the power control variables and noise covariance matrices are optimized via the conventional WMMSE algorithm \cite{Ren:WCL25}.
    \item \textit{Naive fronthaul compression (NaiveFH)}: This method performs independent per-antenna compression with uniform fronthaul rate allocation across antennas by setting $\,\boldsymbol{\Omega}_i =\:$  $\text{diag}(\{\omega_{i,n}\}_{n=1}^{N_A})$ with $\omega_{i,n} = [\sum_{k\in\mathcal{K}}\mathbf{v}_{k,i}\mathbf{v}_{k,i}^H]_{n,n}/\left(2^{C/N_A}-1\right)$ \cite{Liu:Cambridge17,Zhang:arXiv24}. The beamforming vectors are then optimized using the WMMSE algorithm.
    \item \textit{Projected-gradient-descent WMMSE (PGD-WMMSE)}: As a non-CVX first-order baseline, PGD-WMMSE \cite{Pellaco:OJCS22} updates the beamforming vectors using projected-gradient WMMSE steps while determining the fronthaul quantization noise covariance via equal-rate scalar quantization in a manner similar to NaiveFH.
\end{itemize}
The proposed method and all benchmark schemes are initialized as the same randomly generated feasible point. Specifically, for each $i\in\mathcal{M}$, the entries of $\{\tilde{\mathbf{v}}_{k,i}\}_{k\in\mathcal{K}}$ are independently and identically distributed as $\mathcal{CN}(0,1)$, and we construct $\tilde{\boldsymbol{\Omega}}_i\leftarrow \text{diag}(\{\tilde{\omega}_{i,n}\}_{n=1}^{N_A})$, where $\{\tilde{\omega}_{i,n}\}_{n=1}^{N_A}$ are the squared magnitudes of independently generated zero-mean and unit-variance complex Gaussian random variables. The initial $\{\mathbf{v}_{k,i}\}_{k\in\mathcal{K}}$ and $\boldsymbol{\Omega}_i$ are then set to
\begin{align}
    \mathbf{v}_{k,i} \leftarrow \sqrt{c_{F,i}c_{P,i}} \tilde{\mathbf{v}}_{k,i}, \, \forall k\in\mathcal{K}, \text{ and } \boldsymbol{\Omega}_i \leftarrow c_{P,i}\tilde{\boldsymbol{\Omega}}_i,
\end{align}
where $c_{P,i} > 0$ is chosen to satisfy the power constraint \eqref{eq:fully_power_constraints} with equality.
The factor $c_{F,i} \in [0,1]$ is used to ensure that the fronthaul capacity constraint \eqref{eq:fully_fronthaul_constraints} is satisfied. Specifically, we set $c_{F,i} = 1$ if \eqref{eq:fully_fronthaul_constraints} is already satisfied; otherwise, $c_{F,i}$ is determined via bisection so that \eqref{eq:fully_fronthaul_constraints} holds with equality.

\begin{figure}[t]
\centering\!\!\!\!\!\includegraphics[width=0.7\linewidth]{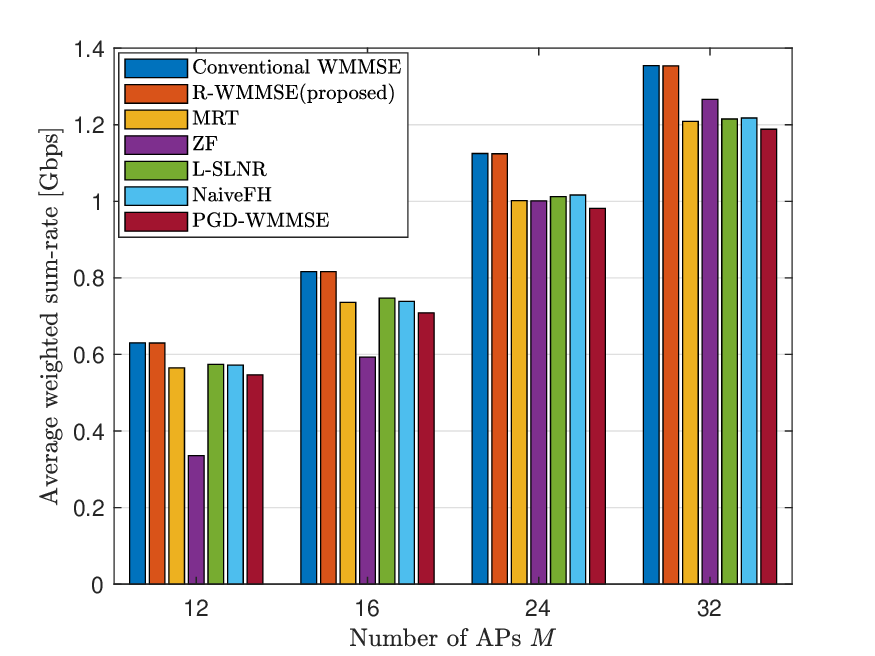}
%\vspace{-5mm}
\caption{\label{fig:avgRsum-vs-M}Average weighted sum-rate versus the number of APs $M$ ($K=20$, $N_A=2$, $C=10$ bps/Hz, $\text{SNR}^{\text{dl}} = 15$ dB, $\text{SNR}^{\text{ul}} = 10$ dB, and $L=16$).}
\end{figure}

\begin{figure}
\centering\!\!\!\!\!\includegraphics[width=0.7\linewidth]{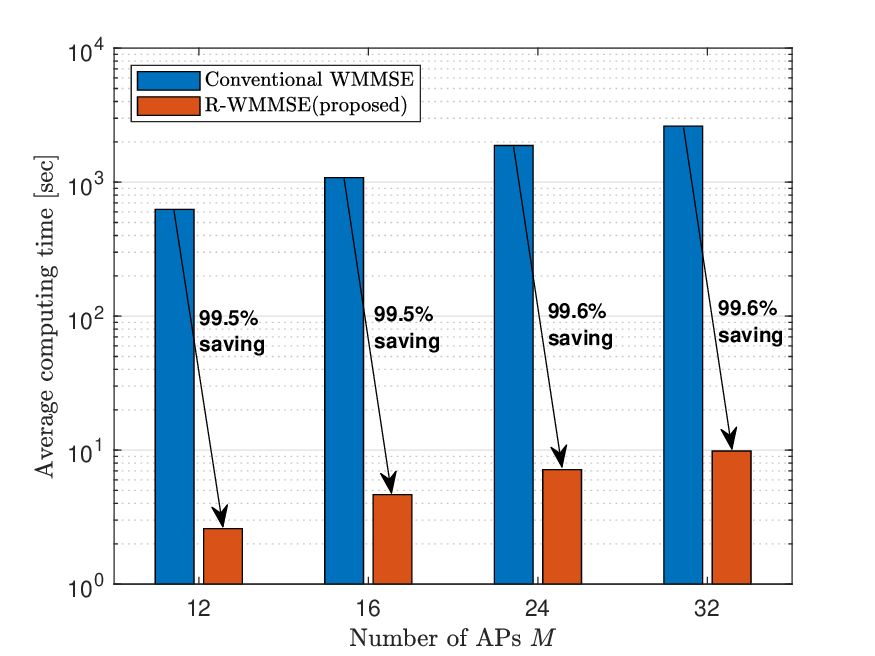}
%\vspace{-5mm}
\caption{\label{fig:avgTime-vs-AP}Average computing time versus the number of APs $M$ ($K=20$, $N_A=2$, $C=10$ bps/Hz, $\text{SNR}^{\text{dl}} = 15$ dB, $\text{SNR}^{\text{ul}} = 10$ dB, and $L=16$).}
\end{figure}

Fig. \ref{fig:avgRsum-vs-M} illustrates how the average weighted sum-rates of various schemes scale with the number of APs $M$ for $K=20$, $N_A=2$, $C=10$ bps/Hz, $\text{SNR}^{\text{dl}} = 15$ dB, $\text{SNR}^{\text{ul}} = 10$ dB, and $L=16$. It is observed that the proposed R-WMMSE scheme achieves weighted sum-rate performance nearly identical to that of the conventional WMMSE scheme across all values of $M$, while requiring significantly reduced computational complexity. Among the baseline schemes, ZF increasingly outperforms MRT in the large-$M$ regime, benefiting from the increased spatial degrees of freedom available for interference suppression.  
In addition, the L-SLNR scheme \cite{Ren:WCL25} consistently provides a slight performance improvement over MRT for all $M$, while still remaining below the conventional WMMSE and proposed R-WMMSE schemes.
The proposed R-WMMSE method outperforms the non-CVX PGD-WMMSE baseline, demonstrating the performance advantage and scalability of the proposed vector-quantization-based design.
Overall, these results confirm that the proposed R-WMMSE remains effective and scalable in massive networks.

Fig. \ref{fig:avgTime-vs-AP} presents the average runtime comparison between the conventional WMMSE and the proposed R-WMMSE algorithms under the same setup as Fig.~\ref{fig:avgRsum-vs-M}.
For all simulated setups, the proposed R-WMMSE scheme reduces the computational time by over 99$\%$ compared to the conventional WMMSE scheme, while achieving the same weighted sum-rate performance as observed in Fig.~\ref{fig:avgRsum-vs-M}.
This complexity reduction is due to the use of closed-form per-AP computations in Algorithm 1, whereas the conventional WMMSE scheme requires solving a convex optimization problem via a general-purpose solver at each iteration.
It should be noted that the runtime comparison in Fig. \ref{fig:avgTime-vs-AP} is influenced by implementation-specific factors and hardware environments, unlike the asymptotic complexity analysis presented in Sec. \ref{sub:complexity}.

\begin{figure}[t]
\centering\!\!\!\!\!\includegraphics[width=0.7\linewidth]{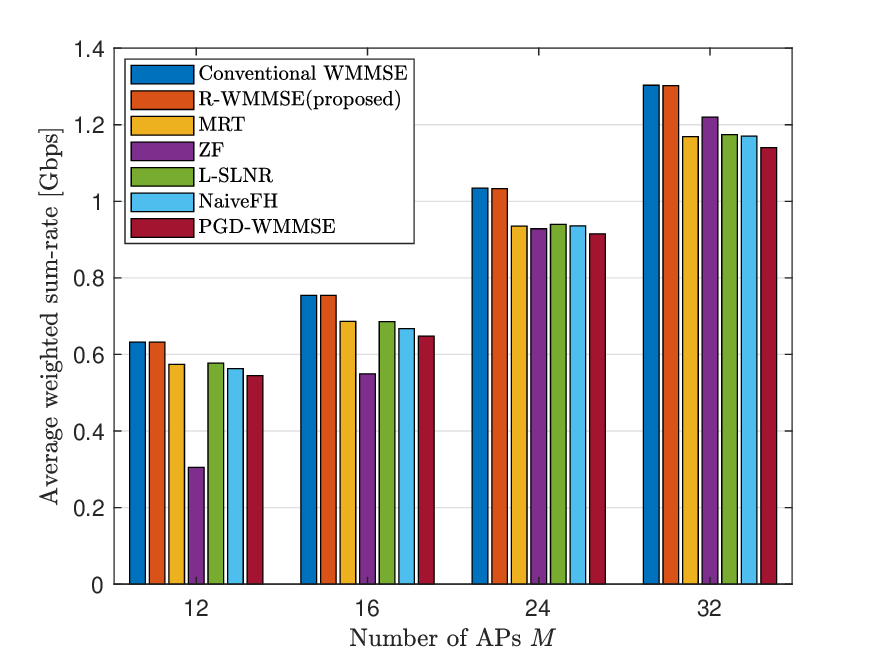}
%\vspace{-5mm}
\caption{\label{fig:avgRsum-vs-M-heteroC}Average weighted sum-rate versus the number of APs $M$ with heterogeneous fronthaul capacities ($K=20$, $N_A=2$, $\text{SNR}^{\text{dl}} = 15$ dB, $\text{SNR}^{\text{ul}} = 10$ dB, and $L=16$).}
\end{figure}

\begin{figure}
\centering\!\!\!\!\!\includegraphics[width=0.7\linewidth]{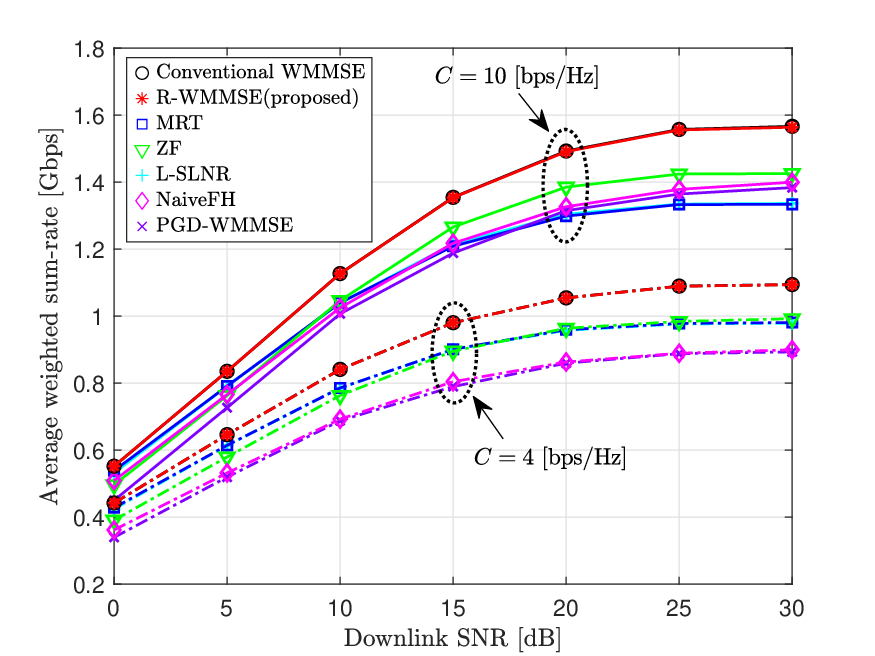}
%\vspace{-5mm}
\caption{\label{fig:avgRsum-vs-SNR}Average weighted sum-rate versus the downlink SNR ($K=20$, $M=32$, $N_A=2$, $C \in \{4, 10\}$ bps/Hz, $\text{SNR}^{\text{ul}} = 10$ dB, and $L=16$).}
\end{figure}

To evaluate the impact of heterogeneous fronthaul capacities, Fig.~\ref{fig:avgRsum-vs-M-heteroC} presents the average weighted sum-rate versus the number of APs $M$, where the fronthaul capacities $\{C_i\}_{i=1}^{M}$ are independently drawn from the uniform distribution $\mathcal{U}(C_{\min},C_{\max})$ with $(C_{\min},C_{\max})=(4,15)$, using the same setup as Fig. \ref{fig:avgRsum-vs-M}. The proposed R-WMMSE algorithm achieves nearly the same weighted sum-rate as the conventional WMMSE algorithm while consistently outperforming the other baseline schemes. These results demonstrate the robustness of the proposed method to heterogeneous fronthaul capacities across APs.

\begin{figure}[t]
\centering\!\!\!\!\!\includegraphics[width=0.7\linewidth]{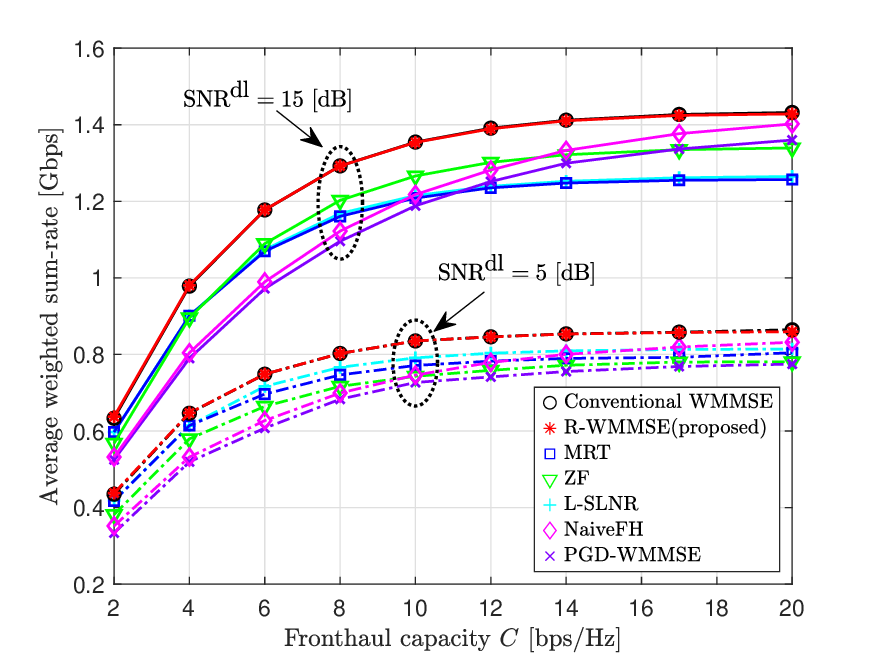}
%\vspace{-5mm}
\caption{\label{fig:avgRsum-vs-CF}Average weighted sum-rate versus the fronthaul capacity $C$ ($K=20$, $M=32$, $N_A=2$, $\text{SNR}^{\text{dl}} \in \{5, 15\}$ dB, $\text{SNR}^{\text{ul}} = 10$ dB, and $L=16$).}
\end{figure}

In Fig. \ref{fig:avgRsum-vs-SNR}, we plot the average weighted sum-rate versus the downlink SNR $\text{SNR}^{\text{dl}}$ for $K=20$, $M=32$, $N_A=2$,  $\text{SNR}^{\text{ul}} = 10$ dB, and $L=16$, with $C \in \{4, 10\}$ bps/Hz. For all simulated $\text{SNR}^{\text{dl}}$ and $C$, the proposed R-WMMSE achieves nearly identical performance to the conventional WMMSE method. 
In addition, it outperforms baseline schemes including the MRT, ZF, L-SLNR, NaiveFH, and PGD-WMMSE. This verifies the effectiveness of the proposed method for optimizing CF-mMIMO networks with reduced computational complexity. It can be observed that the weighted sum-rate performance of all schemes first increases as $\text{SNR}^{\text{dl}}$ increases, then saturates at high SNR. This indicates that the overall system performance of CF-mMIMO is limited by fronthaul capacity even with sufficient downlink transmit power. A larger $C$ yields higher weighted sum-rate performance since the fronthaul quantization distortion is reduced and the precoded signals can be delivered more accurately.

Fig. \ref{fig:avgRsum-vs-CF} depicts the average weighted sum-rate with respect to the fronthaul capacity $C$ for $K=20$, $M=32$, $N_A=2$, $\text{SNR}^{\text{ul}} = 10$ dB, and $L=16$ with $\text{SNR}^{\text{dl}} \in \{5, 15\}$ dB. Similar to Fig. \ref{fig:avgRsum-vs-SNR}, we can see that for all simulated $C$, the proposed R-WMMSE significantly outperforms the MRT, ZF, L-SLNR, NaiveFH, and PGD-WMMSE benchmarks and closely matches the conventional WMMSE, demonstrating its effectiveness. In particular, the comparison with NaiveFH indicates that optimizing the quantization noise covariance matrices provides the most significant benefit in the intermediate $C$ regime. Increasing $C$ improves the weighted sum-rate performance, with the most noticeable gains observed in the low-to-intermediate-$C$ regime. The gains diminish as $C$ becomes sufficiently large because quantization distortion is no longer the dominant limitation and the performance becomes primarily constrained by the wireless downlink conditions, which leads to saturation at large $C$.

\begin{figure}[t]
\centering\!\!\!\!\!\includegraphics[width=0.7\linewidth]{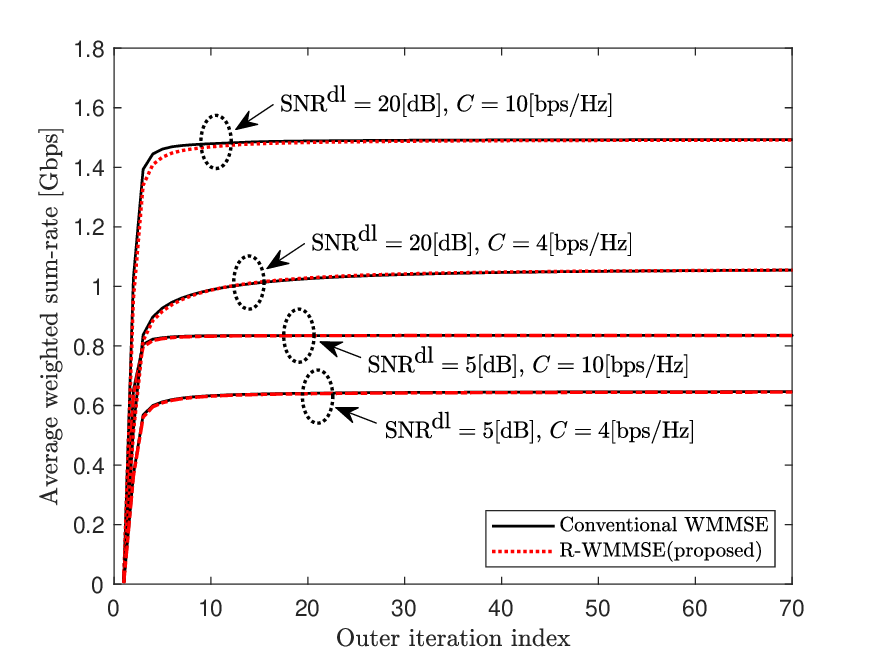}
%\vspace{-5mm}}
\caption{\label{fig:convergence}Average weighted sum-rate versus the outer iteration index ($K=20$, $M=32$, $N_A=2$, $C \in \{4, 10\}$ bps/Hz, $\text{SNR}^{\text{dl}} \in \{5, 20\}$ dB, $\text{SNR}^{\text{ul}} = 10$ dB, and $L=16$).}
\end{figure}

\begin{figure}
\centering\!\!\!\!\!\includegraphics[width=0.7\linewidth]{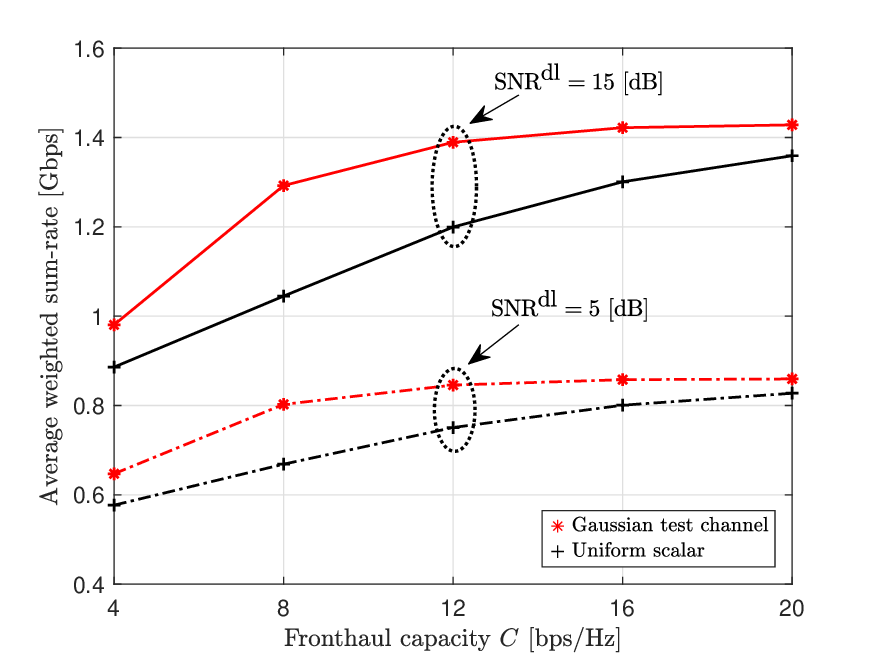}
%\vspace{-5mm}
\caption{\label{fig:comparison-quantization}Average weighted sum-rate versus the fronthaul capacity $C$ ($K=20$, $M=32$, $N_A=2$, $\text{SNR}^{\text{dl}} \in \{5, 15\}$ dB, $\text{SNR}^{\text{ul}} = 10$ dB, and $L=16$).}
\end{figure}

Fig. \ref{fig:convergence} shows the average weighted sum-rate versus the outer iteration index for both the proposed R-WMMSE and the conventional WMMSE schemes under the same setup as Fig.~\ref{fig:avgRsum-vs-SNR}. As shown, the proposed R-WMMSE scheme attains nearly the same convergence speed as the conventional WMMSE, reaching convergence within just a few iterations.

In Fig. \ref{fig:comparison-quantization}, we investigate the impact of the Gaussian test channel model (\ref{eq:fully_beamforming}) for fronthaul quantization.
To this end, we compare the average weighted sum-rates achieved under the Gaussian test channel and uniform scalar quantization for a CF-mMIMO system with $K=20$, $M=32$, $N_A=2$, $\text{SNR}^{\text{dl}} \in \{5, 15\}$ dB, $\text{SNR}^{\text{ul}} = 10$ dB, and $L=16$. 
For uniform scalar quantization, we adopt the additive quantization noise model (AQNM) \cite{Yoo:WCL24}, where the quantized signal is given by a linear combination of the source signal and an uncorrelated quantization noise term, similarly to (\ref{eq:fully_beamforming}) but with larger noise power.
The figure demonstrates that the uniform scalar quantization leads to a noticeable performance degradation compared to the Gaussian test channel. Nevertheless, the performance gap decreases as the fronthaul capacity $C$ increases, since sufficiently small quantization noise powers can be achieved even with uniform scalar quantization when ample fronthaul resources are available. These results show that the Gaussian test channel enables tractable joint beamforming and compression and provides an information-theoretic upper bound under finite fronthaul capacity.

\begin{table}[t]
\caption{Average Weighted Sum-Rate of Non-CVX Methods \\ for Large-Scale Networks}
\label{tab:large_network}
\centering
\renewcommand{\arraystretch}{1.1}
\setlength{\tabcolsep}{6pt}
\begin{tabular}{c|cccc}
\hline\hline
\multirow{2}{*}{\textbf{Schemes}}
& \multicolumn{4}{c}{\textbf{Number of APs $M$}} \\ 
& \textbf{48} & \textbf{64} & \textbf{128} & \textbf{256} \\ \hline
\textbf{R-WMMSE(proposed)} [Gbps]
& 1.7464 & 2.0474 & 2.6116 & 3.2452 \\
PGD-WMMSE [Gbps]
& 1.5226 & 1.7768 & 2.2490 & 2.6586 \\ \hline
Gain [\%] & 14.7 & 15.2 & 16.1 & 22.1 \\ \hline\hline
\end{tabular}
\end{table}

In Table~\ref{tab:large_network}, we evaluate the scalability of the proposed R-WMMSE scheme for large-scale CF-mMIMO networks with up to $M=256$ APs. The simulations are conducted with $K=20$, $N_A=2$, $C=10$ bps/Hz, $\text{SNR}^{\text{dl}}=15$ dB, $\text{SNR}^{\text{ul}}=10$ dB, and $L=16$. To focus on scalable implementations for large network sizes, the comparison includes only non-CVX methods, namely the proposed R-WMMSE and PGD-WMMSE schemes. As shown in Table~\ref{tab:large_network}, the proposed R-WMMSE consistently achieves a higher weighted sum-rate than PGD-WMMSE for all considered values of $M$. In particular, the weighted sum-rate gain of R-WMMSE over PGD-WMMSE increases from $14.7\%$ to $22.1\%$ as $M$ increases from $48$ to $256$, respectively, demonstrating the favorable performance and scalability of the proposed scheme in large-scale CF-mMIMO deployments.

\subsection{With Hybrid Analog-Digital Architectures ($N_R \leq N_A$)} \label{sub:numerical-hybrid-analog-digital}

In this subsection, we evaluate the effectiveness of the proposed R-WMMSE scheme for the hybrid analog-digital architecture described in Sec. \ref{sec:hybrid_extension}. For the analog phase refinement, we set the diminishing rate $r$ to 0.85. As baselines, we consider the following schemes.
\begin{itemize}
    \item \textit{Fully-digital beamforming:} Each AP $i$ is equipped with $N_A=8$ RF chains (i.e., no RF limitation), and only digital beamforming, along with the fronthaul compression, is applied. The digital variables $\{\mathbf{v}, \boldsymbol{\Omega}\}$ are jointly optimized via Algorithm \ref{algorithm1}.
    \item \textit{Random analog beamforming:} The analog beamformers $\mathbf{A}$ have randomly chosen phase values, while the digital variables $\{\mathbf{v}, \boldsymbol{\Omega}\}$ are jointly optimized via Algorithm \ref{algorithm1}.
\end{itemize}
As in the fully-digital case, the proposed method and all benchmarks use the same randomly generated feasible initialization for the digital beamforming vectors and quantization-noise covariance matrices. For the proposed hybrid R-WMMSE method, all analog phase shifts are initialized to zero, i.e., $\mathbf{A}_i=\mathbf{1}$ for all $i$.

\begin{figure}[t]
\centering\!\!\!\!\!\includegraphics[width=0.7\linewidth]{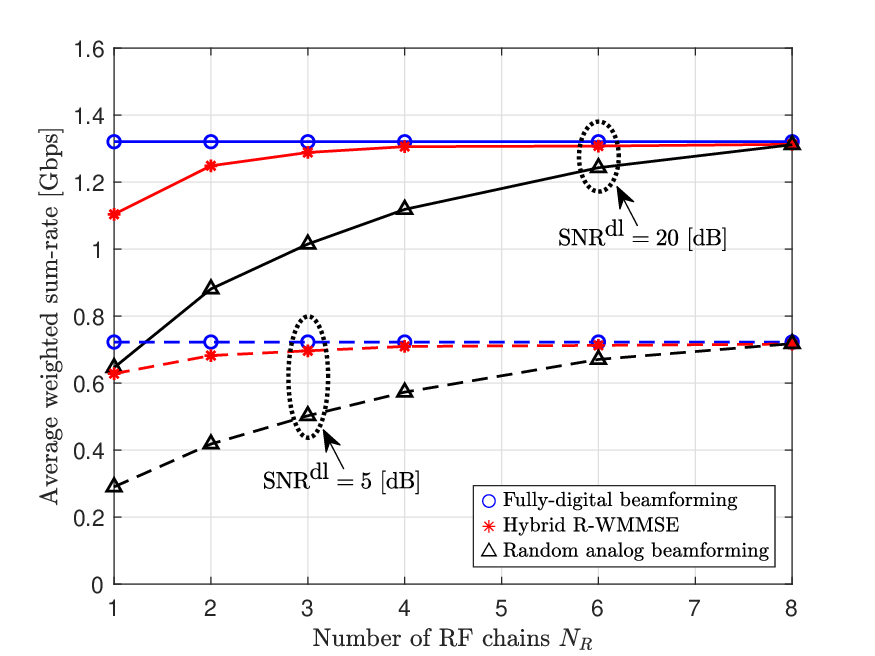}
%\vspace{-5mm}
\caption{\label{fig:hybrid-avgSR-vs-RF}Average weighted sum-rate versus the number of RF chains $N_R$ ($K=16$, $M=20$, $N_A=8$, $C=10$ bps/Hz, $\text{SNR}^{\text{dl}} \in \{5, 20\}$ dB, $\text{SNR}^{\text{ul}} = 10$ dB, and $L=14$).}
\end{figure}

Fig.~\ref{fig:hybrid-avgSR-vs-RF} presents the average weighted sum-rate versus the number of RF chains $N_R$ for $K=16$, $M=20$, $N_A=8$, $C=10$ bps/Hz, $\text{SNR}^{\text{dl}} \in \{5, 20\}$ dB, $\text{SNR}^{\text{ul}} = 10$ dB, and $L=14$. The proposed hybrid R-WMMSE method exhibits superior performance compared with the random analog beamforming baseline when the number of RF chains $N_R$ is small. As $N_R$ increases, the weighted sum-rates achieved by both the hybrid R-WMMSE and the random analog beamforming baseline gradually approach the upper bound performance of fully-digital beamforming with $N_A=N_R=8$. This is because a larger number of RF chains provides more effective spatial degrees of freedom and relaxes the hardware-induced dimensionality constraint of the hybrid architecture. Regardless of $\text{SNR}^{\text{dl}}$, $N_{R}=4$ RF chains are sufficient for the proposed scheme to achieve the upper bound performance of the fully-digital system. These results indicate that the joint design of the analog and digital beamforming is especially crucial when the number of RF chains is limited, whereas the performance gap naturally diminishes as more RF chains become available.

\begin{figure}
\centering\!\!\!\!\!\includegraphics[width=0.7\linewidth]{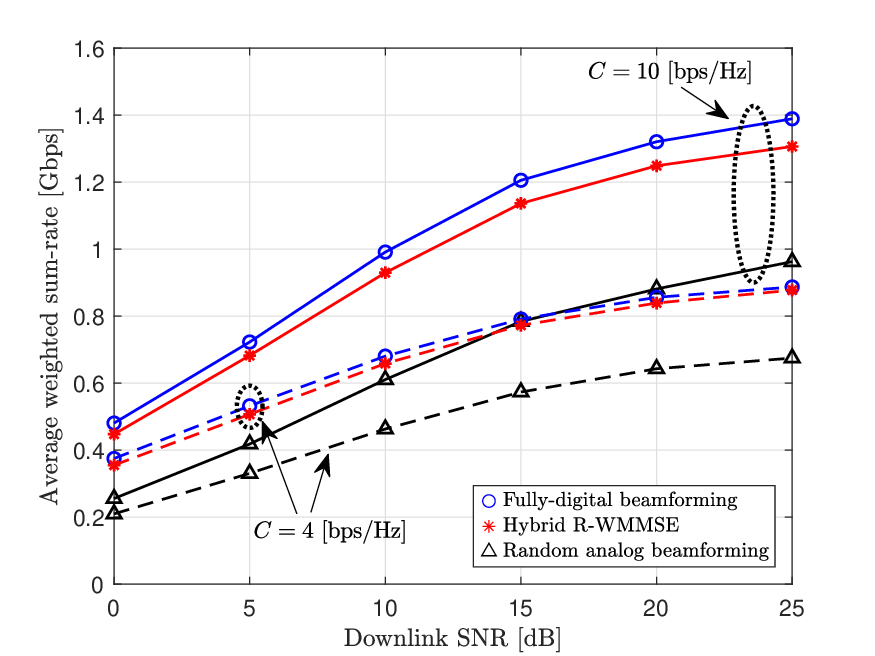}
%\vspace{-5mm}
\caption{\label{fig:hybrid-avgSR-vs-SNR}Average weighted sum-rate versus the downlink SNR ($K=16$, $M=20$, $N_A=8$, $N_R=2$, $C \in \{4, 10\}$ bps/Hz, $\text{SNR}^{\text{ul}} = 10$ dB, and $L=14$).}
\end{figure}

Fig.~\ref{fig:hybrid-avgSR-vs-SNR} shows the average weighted sum-rate versus the downlink SNR in an RF chain-limited hybrid setting with $N_R=2$, using the same schemes as in Fig.~\ref{fig:hybrid-avgSR-vs-RF},  for $K=16$, $M=20$, $N_A=8$, $C \in \{4, 10\}$ bps/Hz, $\text{SNR}^{\text{ul}} = 10$ dB, and $L=14$. As $\text{SNR}^{\text{dl}}$ increases, the average weighted sum-rate increases for all schemes, and a larger fronthaul capacity $C$ yields higher performance due to reduced quantization distortion and more accurate delivery of the precoded signals. A key observation is that, despite the limited number of RF chains $N_R=2$, the proposed hybrid R-WMMSE maintains a small performance gap to the fully-digital benchmark across the entire SNR range. In contrast, the random analog beamforming baseline exhibits noticeably lower performance, and the gap becomes more pronounced as $\text{SNR}^{\text{dl}}$ increases, since the non-optimized analog beamforming cannot fully exploit the available spatial degrees of freedom, especially at higher SNR.
This demonstrates that the proposed joint optimization of the digital and analog beamformers effectively mitigates the performance loss caused by the hybrid precoding structure when the number of RF chains is limited.

\begin{figure}[t]
\centering\!\!\!\!\!\includegraphics[width=0.7\linewidth]{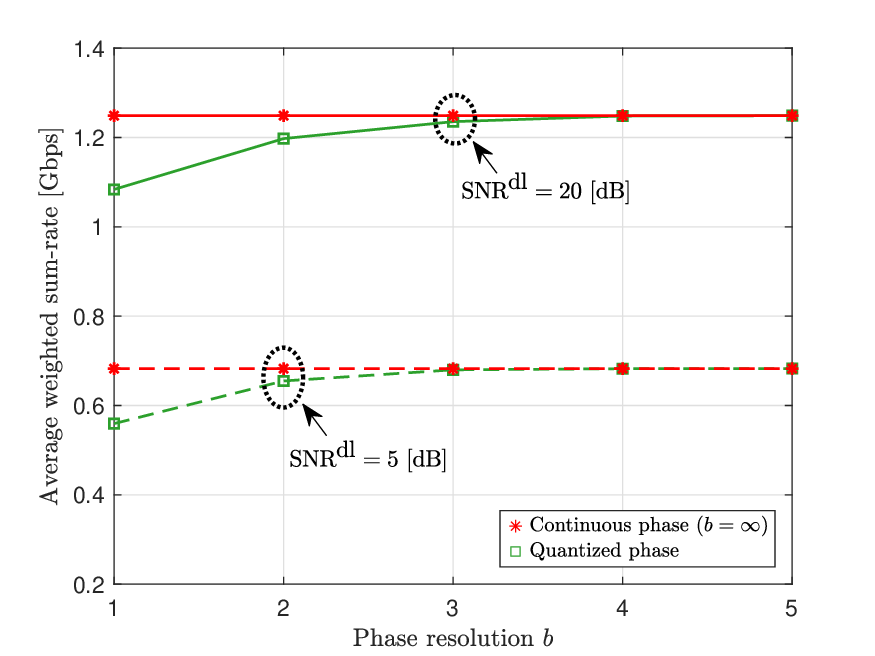}
%\vspace{-5mm}
\caption{\label{fig:hybrid-avgSR-vs-Quanbit}Average weighted sum-rate versus the phase resolution $b$ ($K=16$, $M=20$, $N_A=8$, $N_R=2$, $C=10$ bps/Hz, $\text{SNR}^{\text{dl}} \in \{5, 20\}$ dB, $\text{SNR}^{\text{ul}} = 10$ dB, and $L=14$).}
\end{figure}

In Fig. \ref{fig:hybrid-avgSR-vs-Quanbit}, we evaluate the impact of finite-resolution phase shifts in analog precoders on the performance of the proposed hybrid R-WMMSE scheme.
We adopt a uniform $b$-bit phase-shifter model commonly used in, e.g., \cite{sohrabi:SPAWC15, Chang:TWC24}, where each analog phase is constrained to the discrete set $\theta_{i,n,n'} \in \{2\pi q / 2^b\}_{q=0}^{2^b-1}$.
For comparison, we consider the following two implementations: \textit{continuous phase}, in which the analog beamformer is updated without phase quantization, corresponding to the ideal case $b=\infty$; and \textit{quantized phase}, in which the continuous-phase solution is first obtained by the AO procedure, and phase quantization is applied after convergence. 
Fig.~\ref{fig:hybrid-avgSR-vs-Quanbit} illustrates the average weighted sum-rate versus the phase resolution $b$ for $K=16$, $M=20$, $N_A=8$, $N_R=2$, $C=10$ bps/Hz, $\text{SNR}^{\text{dl}} \in \{5, 20\}$ dB, $\text{SNR}^{\text{ul}} = 10$ dB, and $L=14$.
For both downlink SNR levels, the weighted sum-rate increases monotonically with $b$, as higher resolution allows the quantized phases to better approximate the continuous-phase solution. As $b$ becomes sufficiently large, the performance approaches that of the continuous-phase benchmark, indicating that the quantization-induced loss becomes negligible. 
Overall, these results demonstrate that the proposed hybrid R-WMMSE remains robust to finite-resolution phase shifters and achieves near-optimal performance with a modest number of quantization bits.

\section{Conclusion} \label{sec:conclusion}

We have proposed the R-WMMSE framework for CF-mMIMO systems with finite-capacity fronthaul links, addressing the joint optimization of beamforming and fronthaul compression to maximize the weighted sum-rate under fronthaul capacity constraints.
A sequential per-AP optimization strategy was employed, wherein each subproblem is efficiently solved via Lagrangian duality with closed-form updates.
The proposed framework was further extended to hybrid analog-digital beamforming systems by jointly optimizing the analog beamforming, digital beamforming, and fronthaul compression through an AO approach.
Numerical results validated the effectiveness of the proposed R-WMMSE framework in both fully-digital and hybrid analog-digital setups.

As future work, we plan to develop a deep learning-based method that leverages the optimal structures of beamforming vectors and quantization noise covariance matrices to enable efficient learning.
Moreover, we aim to extend the proposed R-WMMSE framework to integrated sensing and communication (ISAC) systems \cite{Zhu:TWC25} by incorporating an additional sensing metric, as well as to max-min fairness formulations \cite{Zhou:CL20}.
In addition, practical CF-mMIMO systems require efficient optimization algorithms that account for various system constraints. These include user-centric clustering \cite{Lai:TVT25, Mobini:TWC2}, energy and fronthaul efficiency \cite{Kim:TCOM19}, imperfect synchronization \cite{Park:SPL16}, probabilistic fronthaul failures \cite{Lee:TVT19}, and pilot overhead \cite{Xiong:JSTSP}.
Finally, investigating efficient designs and their distributed implementations under alternative functional split options, where beamforming is locally performed at the respective APs, constitutes an interesting direction for future research.

% References
\bibliographystyle{IEEEtran}
\bibliography{REF}

\end{document}